\documentclass{article}

\usepackage[T1]{fontenc}

\usepackage[bitstream-charter]{mathdesign}
\usepackage{amsmath}
\usepackage[scaled=0.92]{PTSans}

\usepackage[
  paper  = letterpaper,
  left   = 1.65in,
  right  = 1.65in,
  top    = 1.0in,
  bottom = 1.0in,
  ]{geometry}
\usepackage{setspace}

\usepackage[usenames,dvipsnames]{xcolor}
\definecolor{shadecolor}{gray}{0.9}

\usepackage[final,expansion=alltext]{microtype}
\usepackage[english]{babel}
\usepackage[parfill]{parskip}
\usepackage{afterpage}
\usepackage{framed}
\usepackage{nicefrac}

\DeclareRobustCommand{\parhead}[1]{\textbf{#1}~}

\usepackage{graphicx}
\usepackage[labelfont=bf]{caption}
\usepackage[format=hang]{subcaption}

\usepackage{booktabs}

\usepackage{natbib}

\usepackage[algoruled]{algorithm2e}

\usepackage{listings}
\lstdefinestyle{mystyle}{
    commentstyle=\color{OliveGreen},
    numberstyle=\tiny\color{black!60},
    stringstyle=\color{BrickRed},
    basicstyle=\ttfamily\scriptsize,
    breakatwhitespace=false,
    breaklines=true,
    captionpos=b,
    keepspaces=true,
    numbers=none,
    numbersep=5pt,
    showspaces=false,
    showstringspaces=false,
    showtabs=false,
    tabsize=2
}
\usepackage{hyperref}
\hypersetup{citecolor=Violet}
\hypersetup{linkcolor=black}
\hypersetup{urlcolor=MidnightBlue}

\usepackage[acronym,smallcaps,nowarn]{glossaries}
\glsdisablehyper{}

\DeclareRobustCommand{\mb}[1]{\ensuremath{\boldsymbol{\mathbf{#1}}}}

\DeclareMathOperator*{\argmax}{arg\,max}
\DeclareMathOperator*{\argmin}{arg\,min}

\DeclareMathOperator*{\diag}{diag}

\newcommand\dif{\mathop{}\!\mathrm{d}}

\newcommand{\bm}{\mathbf{m}}
\newcommand{\bs}{\mathbf{s}}

\newcommand{\bx}{\mathbf{x}}
\newcommand{\by}{\mathbf{y}}
\newcommand{\bz}{\mathbf{z}}

\newcommand{\bmu}{\mb{\mu}}

\newcommand{\btau}{\mb{\tau}}
\newcommand{\bvarphi}{\mb{\varphi}}
\newcommand{\bc}{\mathbf{c}}
\newcommand{\ELBO}{\textsc{elbo}}
\newcommand{\cN}{\mathcal{N}}
\newcommand{\cQ}{\mathcal{Q}}

\newcommand{\g}{\,\vert\,}
\newcommand{\E}{\mathbb{E}}
\newcommand{\EE}[1]{\mathbb{E}\left[#1\right]}
\newcommand{\EEE}[2]{\mathbb{E}_{#1}\left[#2\right]}
\newcommand{\kl}[1]{\textsc{kl}\left(#1\right)}

\newcommand{\realline}{\mathbb{R}}

\newcommand{\mult}{\textrm{Mult}}
\newcommand{\Dir}{\textrm{Dir}}

\newcommand{\const}{\mathrm{const}}

\newcommand{\Gam}{\textrm{Gam}}

\newacronym{KL}{kl}{Kullback-Leibler}
\newacronym{ELBO}{elbo}{evidence lower bound}

\newacronym{EM}{em}{expectation maximization}

\newacronym{MCMC}{mcmc}{Markov chain Monte Carlo}
\newacronym{MC}{mc}{Monte Carlo}

\newacronym{EP}{ep}{expectation propagation}

\newacronym{VI}{vi}{variational inference}
\newacronym{MFVI}{mfvi}{mean-field variational inference}
\newacronym{SVI}{svi}{stochastic variational inference}
\newacronym{CAVI}{cavi}{coordinate ascent variational inference}

\newacronym{AD}{ad}{automatic differentiation}
\newacronym{ADVI}{advi}{automatic differentiation variational inference}
\newacronym{ADVIDIAG}{advi-diag}{automatic differentiation variational
inference diagonal}
\newacronym{ADSVI}{adsvi}{automatic differentiation stochastic variational
inference}

\newacronym{RMSPROP}{rmsprop}{rmsprop}

\newacronym{GMM}{gmm}{Gaussian mixture model}
\newacronym{LDA}{lda}{latent Dirichlet allocation}
\newacronym{ARD}{ard}{automatic relevance determination}

\newacronym{SGA}{sga}{stochastic gradient ascent}

\newacronym{MLE}{mle}{maximum likelihood estimate}

\begin{document}
\title{\textbf{Variational Inference: A Review for Statisticians}}
\author{David M.~Blei\\
  Department of Computer Science and Statistics\\ Columbia
  University\\ \\
  Alp Kucukelbir\\
  Department of Computer Science \\ Columbia University\\ \\
  Jon D.~McAuliffe \\
  Department of Statistics\\University of California, Berkeley}
\maketitle

\bigskip
\begin{abstract}
  One of the core problems of modern statistics is to approximate
  difficult-to-compute probability densities. This problem is especially
  important in Bayesian statistics, which frames all inference about unknown
  quantities as a calculation involving the posterior density. In this
  paper, we review \gls{VI}, a method from machine learning that approximates
  probability densities through optimization. \gls{VI} has been used in many
  applications and tends to be faster than classical methods, such as Markov
  chain Monte Carlo sampling. The idea behind \gls{VI} is to first posit a
  family of densities and then to find the member of that family which is
  close to the target. Closeness is measured by Kullback-Leibler divergence. We
  review the ideas behind mean-field variational inference, discuss the special
  case of \gls{VI} applied to exponential family models, present a full example
  with a Bayesian mixture of Gaussians, and derive a variant that uses
  stochastic optimization to scale up to massive data. We discuss modern
  research in \gls{VI} and highlight important open problems. \gls{VI} is
  powerful, but it is not yet well understood. Our hope in writing this paper
  is to catalyze statistical research on this class of algorithms.
\end{abstract}

\clearpage
\glsresetall{}

\section{Introduction}

One of the core problems of modern statistics is to approximate
difficult-to-compute probability densities.  This problem is
especially important in Bayesian statistics, which frames all
inference about unknown quantities as a calculation about the
posterior.  Modern Bayesian statistics relies on models for which the
posterior is not easy to compute and corresponding algorithms for
approximating them.

In this paper, we review \gls{VI}, a method from machine learning for
approximating probability
densities~\citep{Jordan:1999,wainwright2008graphical}.
Variational inference is widely used to approximate posterior
densities for Bayesian models, an alternative strategy to
\gls{MCMC} sampling.  Compared to \gls{MCMC}, variational inference
tends to be faster and easier to scale to large data---it has been
applied to problems such as large-scale document analysis,
computational neuroscience, and computer vision.  But variational
inference has been studied less rigorously than \gls{MCMC}, and
its statistical properties are less well understood.  In writing this
paper, our hope is to catalyze statistical research on variational
inference.

First, we set up the general problem.  Consider a joint density
of latent variables $\bz = z_{1:m}$ and observations $\bx = x_{1:n}$,
\begin{align*}
  p(\bz, \bx) = p(\bz) p(\bx \g \bz).
\end{align*}
In Bayesian models, the latent variables help govern the distribution
of the data.  A Bayesian model draws the latent variables from a prior
density $p(\bz)$ and then relates them to the observations
through the likelihood $p(\bx \g \bz)$.  Inference in a Bayesian model
amounts to conditioning on data and computing the posterior
$p(\bz \g \bx)$.  In complex Bayesian models, this computation often
requires approximate inference.

For decades, the dominant paradigm for approximate inference has been
\gls{MCMC}~\citep{Hastings:1970,Gelfand:1990}. In \gls{MCMC}, we first
construct an ergodic Markov chain on $\bz$ whose stationary distribution is the
posterior $p(\bz \g \bx)$.  Then, we sample from the chain
to collect samples from the stationary
distribution. Finally, we approximate the posterior with an empirical
estimate constructed from (a subset of) the collected samples.

\gls{MCMC} sampling has evolved into an indispensable tool to the
modern Bayesian statistician. Landmark developments include the
Metropolis-Hastings algorithm \citep{Metropolis:1953,Hastings:1970},
the Gibbs sampler \citep{Geman:1984} and its application to Bayesian
statistics \citep{Gelfand:1990}. \gls{MCMC} algorithms are under
active investigation.  They have been widely studied, extended, and
applied; see \citet{Robert:2004} for a perspective.

However, there are problems for which we cannot easily use this
approach.  These arise particularly when we need an approximate
conditional faster than a simple \gls{MCMC} algorithm can produce,
such as when data sets are large or models are very complex.  In these
settings, variational inference provides a good alternative approach to
approximate Bayesian inference.

Rather than use sampling, the main idea behind variational inference
is to use optimization.  First, we posit a \emph{family} of
approximate densities $\cQ$.
This is a set of densities over the latent variables.
Then, we try
to find the member of that family that minimizes the \gls{KL}
divergence to the exact posterior,
\begin{align}
  \label{eq:general-VI-optimization}
  q^*(\bz) = \argmin_{q(\bz) \in \cQ} \kl{q(\bz) \| p(\bz \g \bx)}.
\end{align}
Finally, we approximate the posterior with the optimized member of the
family $q^*(\cdot)$.

Variational inference thus turns the inference problem into an
optimization problem, and the reach of the family $\cQ$ manages the
complexity of this optimization.  One of the key ideas behind
variational inference is to choose $\cQ$ to be flexible enough to
capture a density close to $p(\bz \g \bx)$, but simple enough for
efficient optimization.\footnote{We focus here on
  $\gls{KL}(q||p)$-based optimization, also called Kullback Leibler
  variational inference~\citep{Barber:2012}.
  \citet{wainwright2008graphical} emphasize that any procedure which
  uses optimization to approximate a density can be termed
  ``variational inference.''  This includes methods like
  expectation propagation~\citep{minka2001expectation}, belief
  propagation~\citep{Yedidia:2001}, or even the Laplace approximation.
  We briefly discuss alternative divergence measures in
  Section\nobreakspace \ref {sec:discussion}.}

We emphasize that \gls{MCMC} and variational inference are different
approaches to solving the same problem.  \gls{MCMC} algorithms sample
a Markov chain; variational algorithms solve an optimization
problem.  \gls{MCMC} algorithms approximate the posterior with samples
from the chain; variational algorithms approximate the posterior with
the result of the optimization.

\parhead{Comparing variational inference and \gls{MCMC}.} When should
a statistician use \gls{MCMC} and when should she use variational
inference?  We will offer some guidance.
\gls{MCMC} methods tend to be more computationally
intensive than variational inference but they also provide guarantees
of producing (asymptotically) exact samples from the target density
\citep{Robert:2004}.  Variational inference does not enjoy
such guarantees---it can only find a density close to the
target---but tends to be faster than \gls{MCMC}.  Because it rests on
optimization, variational inference easily takes advantage of methods
like stochastic optimization~\citep{Robbins:1951,Kushner:1997} and
distributed optimization (though some \gls{MCMC} methods can also
exploit these innovations~\citep{Welling:2011,Ahmed:2012}).

Thus, variational inference is suited to large data
sets and scenarios where we want to quickly explore many models;
\gls{MCMC} is suited to smaller data sets and scenarios where we
happily pay a heavier computational cost for more precise samples.
For example, we might use \gls{MCMC} in a setting where we spent 20
years collecting a small but expensive data set, where we
are confident that our model is appropriate, and where we
require precise inferences.  We
might use variational inference when fitting a probabilistic model of
text to one billion text documents and where the inferences will be
used to serve search results to a large population of users.  In this
scenario, we can use distributed computation and stochastic
optimization to scale and speed up inference, and we can easily
explore many different models of the data.

Data set size is not the only consideration. Another factor is the
geometry of the posterior distribution. For example, the posterior
of a mixture model admits multiple modes, each corresponding label
permutations of the components. Gibbs sampling, if the model
permits, is a powerful approach to sampling from such target
distributions; it quickly focuses on one of the modes. For mixture
models where Gibbs sampling is not an option, variational inference
may perform better than a more general \gls{MCMC} technique
(e.g.,~Hamiltonian Monte Carlo), even for small datasets
\citep{kucukelbir2015automatic}. Exploring the interplay between
model complexity and inference (and between variational inference
and \gls{MCMC}) is an exciting avenue for future research (see
Section\nobreakspace \ref {sec:open-problems}).

The relative accuracy of variational inference and
\gls{MCMC} is still unknown.  We do know that variational
inference generally underestimates the variance of the posterior density;
this is a consequence of its objective function.  But, depending on
the task at hand, underestimating the variance may be acceptable.
Several lines of empirical research have shown that variational
inference does not necessarily suffer in accuracy, e.g., in terms of
posterior predictive
densities~\citep{Blei:2006e,Braun:2010,kucukelbir2016automatic};
other research focuses on where variational inference falls short,
especially around the posterior variance, and tries to more closely
match the inferences made by \gls{MCMC} \citep{giordano2015linear}.  In
general, a statistical theory and understanding around variational
inference is an important open area of research (see
Section\nobreakspace \ref {sec:theory}).  We can envision future results that outline which
classes of models are particularly suited to each algorithm and
perhaps even theory that bounds their accuracy.  More broadly,
variational inference is a valuable tool, alongside \gls{MCMC}, in the
statistician's toolbox.

It might appear to the reader that variational inference is only
relevant to Bayesian analysis.  Indeed, both variational inference and
\gls{MCMC} have had a significant impact on applied Bayesian
computation and we will be focusing on Bayesian models here. We
emphasize, however, that these techniques also apply more generally to
computation about intractable densities.  \gls{MCMC} is a tool for
simulating from densities and variational inference is a tool for
approximating densities.  One need not be a Bayesian to have use for
variational inference.

\parhead{Research on variational inference.}  The development of
variational techniques for Bayesian inference followed two parallel,
yet separate, tracks. \citet{Peterson:1987} is arguably the first
variational procedure for a particular model: a neural network. This
paper, along with insights from statistical mechanics
\citep{parisi1988statistical}, led to a flurry of variational
inference procedures for a wide class of models
\citep{saul1996mean,jaakkola1996computing,jaakkola1997variational,
  ghahramani1997factorial,Jordan:1999}.  In parallel,
\citet{Hinton:1993} proposed a variational algorithm for a similar
neural network model. \citet{Neal:1999} (first published in 1993) made
important connections to the \gls{EM}
algorithm~\citep{Dempster:1977}, which then led to a variety of
variational inference algorithms for other types of models
\citep{Waterhouse:1996,mackay1997ensemble}.

Modern research on variational inference focuses on several aspects:
tackling Bayesian inference problems that involve massive
data; using improved optimization methods for solving
Equation\nobreakspace \textup {(\ref {eq:general-VI-optimization})} (which is usually subject to local
minima); developing generic variational inference, algorithms that are
easy to apply to a wide class of models; and increasing the accuracy
of variational inference, e.g., by stretching the boundaries of
$\mathcal{Q}$ while managing complexity in optimization.

\parhead{Organization of this paper.}  Section\nobreakspace \ref {sec:vi} describes the
basic ideas behind the simplest approach to variational inference:
mean-field inference and coordinate-ascent optimization.
Section\nobreakspace \ref {sec:mog} works out the details for a Bayesian mixture of
Gaussians, an example model familiar to many readers.
Sections\nobreakspace \ref {sec:exp-fam} and\nobreakspace  \ref {sec:cond-conj} describe variational inference
for the class of models where the joint density of the latent and
observed variables are in the exponential family---this includes many
intractable models from modern Bayesian statistics and reveals deep
connections between variational inference and the Gibbs sampler
of~\citet{Gelfand:1990}.  Section\nobreakspace \ref {sec:svi} expands on this algorithm to
describe stochastic variational inference~\citep{Hoffman:2013}, which
scales variational inference to massive data using stochastic
optimization~\citep{Robbins:1951}. Finally, with these foundations in
place, Section\nobreakspace \ref {sec:discussion} gives a perspective on the
field---applications in the research literature, a survey of
theoretical results, and an overview of some open problems.

\section{Variational inference}
\label{sec:vi}

The goal of variational inference is to approximate a conditional
density of latent variables given observed variables. The key
idea is to solve this problem with optimization. We use a family of
densities over the latent variables, parameterized by free
``variational parameters.'' The optimization finds the member of this
family, i.e., the setting of the parameters, that is closest in
\gls{KL} divergence to the conditional of interest. The fitted
variational density then serves as a proxy for the exact
conditional density. (All vectors defined below are column vectors, unless
stated otherwise.)

\subsection{The problem of approximate inference}
\label{sec:latent-var-models}

Let $\bx = x_{1:n}$ be a set of observed variables and $\bz = z_{1:m}$
be a set of latent variables, with joint density $p(\bz,
\bx)$. We omit constants, such as hyperparameters, from the notation.

The inference problem is to compute the conditional density of
the latent variables given the observations, $p(\bz \g \bx)$. This
conditional can be used to produce point or interval estimates of the
latent variables, form predictive densities of new data, and more.

We can write the conditional density as
\begin{align}
  p(\bz \g \bx) = \frac{p(\bz,  \bx)}{p(\bx)}.
\end{align}
The denominator contains the marginal density of the observations, also
called the \emph{evidence}. We calculate it by marginalizing out the
latent variables from the joint density,
\begin{align}
  p(\bx) = \int\! p(\bz, \bx) \dif \bz.
  \label{eq:marginal}
\end{align}
For many models, this evidence integral is unavailable in closed form
or requires exponential time to compute. The evidence is what we need
to compute the conditional from the joint; this is why inference in
such models is hard.

Note we assume that all unknown quantities of interest are represented
as latent random variables.  This includes parameters that
might govern all the data, as found in Bayesian models, and latent
variables that are ``local'' to individual data points.

\parhead{Bayesian mixture of Gaussians.} Consider a Bayesian mixture
of unit-variance univariate Gaussians. There are $K$ mixture components,
corresponding to $K$ Gaussian distributions with means
$\bmu = \{\mu_1, \ldots, \mu_K\}$. The mean parameters are drawn
independently from a common prior $p(\mu_k)$, which we assume to be a
Gaussian $\cN(0, \sigma^2)$; the prior variance $\sigma^2$ is a
hyperparameter. To generate an observation $x_i$ from the model, we
first choose a cluster assignment $c_i$.  It
indicates which latent cluster $x_i$ comes from and is drawn from a categorical
distribution over $\{1, \dots, K\}$.  (We encode $c_i$ as
an indicator $K$-vector, all zeros except for a one in the position
corresponding to $x_i$'s cluster.)
We then draw $x_i$ from the corresponding Gaussian $\cN (c_i^\top \bmu, 1)$.

The full hierarchical model is
\begin{align}
  \mu_k & \sim \cN(0, \sigma^2),
         &k &= 1, \ldots, K, \\
  c_i & \sim \textrm{Categorical}(\nicefrac{1}{K}, \ldots, \nicefrac{1}{K}),
         &i &= 1, \ldots , n, \\
  x_i \g c_i, \bmu & \sim \cN\left(c_i^\top \bmu, 1\right)
         &i &= 1, \ldots, n.
\end{align}
For a sample of size $n$, the joint density of latent and
observed variables is
\begin{align}
  p(\bmu, \bc, \bx) =
  p(\bmu) \prod_{i=1}^{n} p(c_i) p(x_i \g c_i, \bmu).
  \label{eq:gmm}
\end{align}
The latent variables are $\bz = \{\bmu, \bc\}$, the $K$ class means
and $n$ class assignments.

Here, the evidence is
\begin{align}
  p(\bx) = \int\! p(\bmu) \prod_{i=1}^{n} \sum_{c_i}
  p(c_i) p(x_i \g c_i, \bmu) \dif{\bmu}.
  \label{eq:gmm-marginal}
\end{align}
The integrand in~Equation\nobreakspace \textup {(\ref {eq:gmm-marginal})} does not contain a separate
factor for each $\mu_k$. (Indeed, each $\mu_k$ appears in all $n$
factors of the integrand.) Thus, the integral
in~Equation\nobreakspace \textup {(\ref {eq:gmm-marginal})} does not reduce to a product of
one-dimensional integrals over the $\mu_k$'s. The time complexity of numerically
evaluating the $K$-dimensional integral is $\mathcal{O}(K^n)$.

If we distribute the product over the sum in~\eqref{eq:gmm-marginal} and
rearrange, we can write the evidence as a sum over all possible
configurations $\bc$ of cluster assignments,
\begin{align}
  p(\bx)
  &=
  \sum_{\bc} p(\bc) \int\! p(\bmu) \prod_{i=1}^{n} p(x_i \g c_i, \bmu)
  \dif{\bmu}.
\end{align}
Here each individual integral is computable, thanks to the conjugacy
between the Gaussian prior on the components and the Gaussian
likelihood. But there are $K^n$ of them, one for each configuration of
the cluster assignments. Computing the evidence remains
exponential in $K$, hence intractable.

\subsection{The evidence lower bound}
\label{sec:elbo}

In variational inference, we specify a family $\mathcal{Q}$ of
densities over the latent variables. Each $q(\bz) \in \mathcal{Q}$
is a candidate approximation to the exact conditional. Our goal is to
find the best candidate, the one closest in \gls{KL} divergence to the
exact conditional.\footnote{ The \gls{KL} divergence is an
  information-theoretic measure of proximity between two
  densities. It is asymmetric---that is,
  $\kl{q \| p} \neq \kl{p \| q}$---and nonnegative. It is minimized
  when $q(\cdot) = p(\cdot)$.}  Inference now amounts to solving the
following optimization problem,
\begin{align}
  q^*(\bz) = \argmin_{q(\bz) \in \mathcal{Q}} \kl{q(\bz) \| p(\bz \g
  \bx)}.
  \label{eq:variational-optimization}
\end{align}
Once found, $q^*(\cdot)$ is the best approximation of the conditional,
within the family $\cQ$.  The complexity of the family determines the
complexity of this optimization.

However, this objective is not computable because it requires
computing the evidence $\log p(\bx)$ in Equation\nobreakspace \textup {(\ref {eq:marginal})}.  (That the
evidence is hard to compute is why we appeal to approximate inference
in the first place.) To see why, recall that \gls{KL} divergence is
\begin{align}
  \kl{q(\bz) \| p(\bz \g \bx)} = \EE{\log
    q(\bz)} - \EE{\log p(\bz \g \bx)},
  \label{eq:kl-simple}
\end{align}
where all expectations are taken with respect to $q(\bz)$.  Expand the
conditional,
\begin{align}
  \kl{q(\bz) \| p(\bz \g \bx)} =
  \E\left[\log q(\bz)\right] -
  \E\left[\log p(\bz, \bx)\right] +
  \log p(\bx).  \label{eq:kl}
\end{align}
This reveals its dependence on $\log p(\bx)$.

Because we cannot compute the \gls{KL}, we optimize an alternative
objective that is equivalent to the \gls{KL} up to an added constant,
\begin{align}
  \label{eq:elbo}
  \ELBO(q) =
  \EE{\log p(\bz, \bx)} -
  \EE{\log q(\bz)}.
\end{align}
This function is called the \gls{ELBO}.  The \gls{ELBO} is the
negative \gls{KL} divergence of Equation\nobreakspace \textup {(\ref {eq:kl})} plus $\log p(\bx)$, which
is a constant with respect to $q(\bz)$.  Maximizing the \gls{ELBO} is
equivalent to minimizing the \gls{KL} divergence.

Examining the \gls{ELBO} gives intuitions about the optimal
variational density.  We rewrite the \gls{ELBO} as a sum of the
expected log likelihood of the data and the \gls{KL} divergence
between the prior $p(\bz)$ and $q(\bz)$,
\begin{align*}
  \ELBO(q) &= \EE{\log p(\bz)} + \EE{\log p(\bx \g \bz)} - \EE{\log
    q(\bz)} \\
  &= \EE{\log p(\bx \g \bz)} - \kl{q(\bz) \| p(\bz)}.
\end{align*}
Which values of $\bz$ will this objective encourage $q(\bz)$ to place
its mass on? The first term is an expected likelihood; it encourages
densities that place their mass on configurations of the latent
variables that explain the observed data. The second term is the
negative divergence between the variational density and the
prior; it encourages densities close to the prior.  Thus the
variational objective mirrors the usual balance between likelihood and
prior.

Another property of the \gls{ELBO} is that it lower-bounds the (log)
evidence, $\log p(\bx) \geq \ELBO(q)$ for any $q(\bz)$. This explains
the name. To see this notice that Equations\nobreakspace \textup {(\ref {eq:kl})} and\nobreakspace  \textup {(\ref {eq:elbo})} give the
following expression of the evidence,
\begin{align}
  \log p(\bx) = \kl{q(\bz) \| p(\bz \g \bx)} +
  \ELBO(q).
\end{align}
The bound then follows from the fact that
$\kl{\cdot} \geq 0$~\citep{Kullback:1951}.  In the original literature
on variational inference, this was derived through Jensen's
inequality~\citep{Jordan:1999}.

The relationship between the \gls{ELBO} and $\log p(\bx)$ has led to
using the variational bound as a model selection criterion.  This has
been explored for mixture models \citep{Ueda:2002,McGrory:2007}
and more generally~\citep{Beal:2003a}.  The premise is that the bound
is a good approximation of the marginal likelihood, which provides a
basis for selecting a model.  Though this sometimes works in practice,
selecting based on a bound is not justified in theory.
Other research has used
variational approximations in the log predictive density to use
\gls{VI} in cross-validation based model selection~\citep{Nott:2012}.

Finally, many readers will notice that the first term of the
\gls{ELBO} in Equation\nobreakspace \textup {(\ref {eq:elbo})} is the expected complete log-likelihood,
which is optimized by the \gls{EM}
algorithm~\citep{Dempster:1977}. The \gls{EM} algorithm was designed
for finding maximum likelihood estimates in models with latent
variables. It uses the fact that the \gls{ELBO} is equal to the log
likelihood $\log p(\bx)$ (i.e., the log evidence) when
$q(\bz) = p(\bz \g \bx)$.  \gls{EM} alternates between computing the
expected complete log likelihood according to $p(\bz \g \bx)$ (the E
step) and optimizing it with respect to the model parameters (the M
step). Unlike variational inference, \gls{EM} assumes the expectation
under $p(\bz \g \bx)$ is computable and uses it in otherwise difficult
parameter estimation problems. Unlike \gls{EM}, variational inference
does not estimate fixed model parameters---it is often used in a
Bayesian setting where classical parameters are treated as latent
variables. Variational inference applies to models where we cannot
compute the exact conditional of the latent variables.\footnote{Two
  notes: (a) Variational \gls{EM} is the \gls{EM} algorithm with a
  variational E-step, i.e., a computation of an approximate
  conditional. (b) The coordinate ascent algorithm of Section\nobreakspace \ref {sec:cavi}
  can look like the \gls{EM} algorithm. The ``E step'' computes
  approximate conditionals of local latent variables; the ``M step''
  computes a conditional of the global latent variables.  }

\subsection{The mean-field variational family}
\label{sec:mff}

We described the \gls{ELBO}, the variational objective function in the
optimization of Equation\nobreakspace \textup {(\ref {eq:variational-optimization})}. We now describe a
variational family $\cQ$, to complete the specification of the
optimization problem.  The complexity of the family determines the
complexity of the optimization; it is more difficult to optimize over
a complex family than a simple family.

In this review we focus on the \emph{mean-field variational family},
where the latent variables are mutually independent and each governed
by a distinct factor in the variational density.  A generic member
of the mean-field variational family is
\begin{align}
  \label{eq:mf-family}
  q(\bz) = \prod_{j=1}^{m} q_j(z_j).
\end{align}
Each latent variable $z_j$ is governed by its own variational factor,
the density
$q_j(z_j)$.  In optimization, these variational factors are chosen to
maximize the \gls{ELBO} of~Equation\nobreakspace \textup {(\ref {eq:elbo})}.

We emphasize that the variational family is not a model of the
observed data---indeed, the data $\bx$ does not appear in
Equation\nobreakspace \textup {(\ref {eq:mf-family})}. Instead, it is the \gls{ELBO}, and the
corresponding \gls{KL} minimization problem, that connects the fitted
variational density to the data and model.

Notice we have not specified the parametric form of the individual
variational factors. In principle, each can take on any parametric
form appropriate to the corresponding random variable. For example, a
continuous variable might have a Gaussian factor; a categorical
variable will typically have a categorical factor. We will see in
Sections\nobreakspace \ref {sec:ef},  \ref {sec:exp-fam} and\nobreakspace  \ref {sec:cond-conj} that there are many models for which
properties of the model determine optimal forms of the mean-field
variational factors $q_j(z_j)$.

Finally, though we focus on mean-field inference in this review,
researchers have also studied more complex families.  One way to
expand the family is to add dependencies between the
variables~\citep{Saul:1996a,Barber:1999a}; this is called structured
variational inference.  Another way to expand the family is to
consider mixtures of variational densities, i.e., additional
latent variables within the variational family~\citep{Lawrence:1998}.
Both of these methods potentially improve the fidelity of the
approximation, but there is a trade off. Structured and mixture-based
variational families come with a more difficult-to-solve variational
optimization problem.

\parhead{Bayesian mixture of Gaussians (continued).}  Consider again
the Bayesian mixture of Gaussians. The mean-field variational family
contains approximate posterior densities of the form
\begin{align}
  q(\bmu, \bc) = \prod_{k=1}^{K} q(\mu_k; m_k, s^2_k)
  \prod_{i=1}^{n} q(c_i; \varphi_i).
  \label{eq:gmm-mf-family}
\end{align}
Following the mean-field recipe, each latent variable is governed by
its own variational factor.  The factor
$q(\mu_k; m_k, s^2_k)$ is a Gaussian distribution
on the $k$th mixture component's mean parameter; its mean is
$m_k$ and its variance is $s^2_k$.  The factor
$q(c_i; \varphi_i)$ is a distribution on the $i$th observation's
mixture assignment; its assignment probabilities are a $K$-vector
$\varphi_i$.

Here we have asserted parametric forms for these factors: the mixture
components are Gaussian with variational parameters (mean and
variance) specific to the $k$th cluster; the cluster assignments are
categorical with variational parameters (cluster probabilities)
specific to the $i$th data point.  In fact, these are the optimal
forms of the mean-field variational density for the mixture of
Gaussians.

With the variational family in place, we have completely specified the
variational inference problem for the mixture of Gaussians.  The
\gls{ELBO} is defined by the model definition in Equation\nobreakspace \textup {(\ref {eq:gmm})} and the
mean-field family in Equation\nobreakspace \textup {(\ref {eq:gmm-mf-family})}.  The corresponding
variational optimization problem maximizes the \gls{ELBO} with respect
to the variational parameters, i.e., the Gaussian parameters for each
mixture component and the categorical parameters for each cluster
assignment.  We will see this example through in Section\nobreakspace \ref {sec:mog}.

\parhead{Visualizing the mean-field approximation.}  The mean-field family is
expressive because it can capture any marginal density of the
latent variables.  However, it cannot capture correlation between
them.  Seeing this in action reveals some of the intuitions and
limitations of mean-field variational inference.

\begin{figure}[t]
  \centering
  \includegraphics[width=3in]{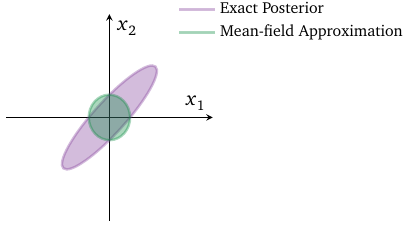}
  \caption{Visualizing the mean-field approximation to a two-dimensional
  Gaussian posterior. The ellipses show the effect of mean-field factorization.
  (The ellipses are $2 \sigma$ contours of the Gaussian distributions.)}
  \label{fig:accuracy}
\end{figure}

Consider a two dimensional Gaussian distribution, shown in violet in
Figure \ref{fig:accuracy}. This density is highly correlated, which
defines its elongated shape.

The optimal mean-field variational approximation to this posterior is
a product of two Gaussian distributions. Figure\nobreakspace \ref {fig:accuracy} shows the
mean-field variational density after maximizing the \gls{ELBO}. While
the variational approximation has the same mean as the original
density, its covariance structure is, by construction, decoupled.

Further, the marginal variances of the approximation under-represent
those of the target density. This is a common effect in
mean-field variational inference and, with this example, we can see
why.  The \gls{KL} divergence from the approximation to the posterior
is in Equation\nobreakspace \textup {(\ref {eq:kl-simple})}.  It penalizes placing mass in $q(\cdot)$ on
areas where $p(\cdot)$ has little mass, but penalizes less the
reverse.  In this example, in order to successfully match the marginal
variances, the circular $q(\cdot)$ would have to expand into territory
where $p(\cdot)$ has little mass.

\subsection{Coordinate ascent mean-field variational inference}
\label{sec:cavi}

Using the \gls{ELBO} and the mean-field family, we have cast
approximate conditional inference as an optimization problem.  In this
section, we describe one of the most commonly used algorithms for
solving this optimization problem, \gls{CAVI}~\citep{Bishop:2006}. \gls{CAVI}
iteratively optimizes each
factor of the mean-field variational density, while holding the
others fixed.  It climbs the \gls{ELBO} to a local optimum.

\parhead{The algorithm.}  We first state a result.  Consider the
$j$th latent variable $z_j$.  The \emph{complete conditional} of
$z_j$ is its conditional density given all of the other latent
variables in the model and the observations,
$p(z_j \g \bz_{-j}, \bx)$.  Fix the other variational
factors $q_{\ell}(z_{\ell})$, $\ell \neq j$.  The optimal
$q_{j}(z_j)$ is then proportional to the exponentiated expected log of the
complete conditional,
\begin{align}
  q^*_{j}(z_j) \propto
  \exp\left\{
    \E_{-j}\left[
      \log p(z_j \g \bz_{-j}, \bx)
    \right]
  \right\}.
  \label{eq:qstar}
\end{align}
The expectation in Equation\nobreakspace \textup {(\ref {eq:qstar})} is with respect to the (currently
fixed) variational density over $\bz_{-j}$, that is,
$\prod_{\ell \neq j} q_\ell(z_\ell)$. Equivalently, Equation\nobreakspace \textup {(\ref {eq:qstar})} is
proportional to the exponentiated log of the joint,
\begin{align}
  q_j^*(z_j) \propto
  \exp\left\{
    \E_{-j}\left[
      \log p(z_j, \bz_{-j}, \bx)
    \right]
  \right\}.
  \label{eq:optimal-qj}
\end{align}
Because of the mean-field family assumption---that all the latent variables are
independent---the expectations on the right hand side do not involve
the $j$th variational factor.  Thus this is a valid coordinate update.

These equations underlie the \gls{CAVI} algorithm, presented as
Algorithm\nobreakspace \ref {alg:cavi}. We maintain a set of variational factors
$q_{\ell}(z_{\ell})$. We iterate through them, updating $q_{j}(z_j)$ using
Equation\nobreakspace \textup {(\ref {eq:optimal-qj})}. \gls{CAVI} goes uphill on the \gls{ELBO} of
Equation\nobreakspace \textup {(\ref {eq:elbo})}, eventually finding a local optimum.
As examples we show \gls{CAVI} for a mixture of Gaussians in Section\nobreakspace \ref {sec:mog} and
for a nonconjugate linear regression in Appendix\nobreakspace \ref {app:bayes_linear_regression}.

\gls{CAVI} can also be seen as a ``message passing''
algorithm~\citep{Winn:2005}, iteratively updating each random
variable's variational parameters based on the variational parameters
of the variables in its Markov blanket. This perspective enabled
the design of
automated software for a large class of
models~\citep{Wand:2011,Minka:2014}.  Variational message passing
connects variational inference to the classical theories of
graphical models and probabilistic
inference~\citep{Pearl:1988,Lauritzen:1988}.  It has been extended to
nonconjugate models~\citep{Knowles:2011} and generalized via factor
graphs~\citep{Minka:2005}.

Finally, \gls{CAVI} is closely related to Gibbs
sampling~\citep{Geman:1984,Gelfand:1990}, the classical workhorse of
approximate inference.  The Gibbs sampler maintains a realization of
the latent variables and iteratively samples from each variable's
complete conditional.  Equation\nobreakspace \textup {(\ref {eq:optimal-qj})} uses the same complete
conditional. It takes the expected log, and uses this quantity to
iteratively set each variable's variational factor.\footnote{Many
  readers will know that we can significantly speed up the Gibbs
  sampler by marginalizing out some of the latent variables; this is
  called collapsed Gibbs sampling. We can speed up variational
  inference with similar reasoning; this is called collapsed
  variational inference. It has been developed for the same class of
  models described here~\citep{Sung:2008,Hensman:2012a}.
  These ideas are outside the scope of our review.}

\glsreset{CAVI}
\begin{algorithm}[t]
\setstretch{1.25}
\KwIn{A model $p(\bx, \bz)$, a data set $\bx$}
\KwOut{A variational density $q(\bz) = \prod_{j=1}^{m} q_{j}(z_j)$}
\textbf{Initialize:} Variational factors $q_{j}(z_j)$ \\
\While{the \gls{ELBO} has not converged} {
  \For{$j \in \{1, \ldots, m\}$} {
    Set $q_{j}(z_j) \propto \exp\{\E_{-j}[\log p(z_j \g \bz_{-j}, \bx)]\}$\\
  }
  Compute $\ELBO(q) = \EE{\log p(\bz, \bx)} - \EE{\log q(\bz)}$
}
\Return{$q(\bz)$}
\caption{\Gls{CAVI}}
\label{alg:cavi}
\end{algorithm}

\parhead{Derivation.} We now derive the coordinate update in
Equation\nobreakspace \textup {(\ref {eq:optimal-qj})}. The idea appears in~\citet{Bishop:2006}, but the
argument there uses gradients, which we do not. Rewrite the \gls{ELBO} of
Equation\nobreakspace \textup {(\ref {eq:elbo})} as a function of the $j$th variational factor $q_j(z_j)$,
absorbing into a constant the terms that do not depend on it,
\begin{align}
  \label{eq:coordinate-elbo}
  \ELBO(q_j) =
  \E_{j}\left[ \E_{{-j}}\left[\log p(z_j, \bz_{-j}, \bx) \right] \right]
  - \E_{j}\left[\log q_{j}(z_j)\right] + \const.
\end{align}
We have rewritten the first term of the \gls{ELBO} using iterated expectation.
The second term we have decomposed, using the independence of the variables
(i.e., the mean-field assumption) and retaining only the term that depends
on $q_j(z_j)$.

Up to an added constant, the objective function in
Equation\nobreakspace \textup {(\ref {eq:coordinate-elbo})} is equal to the negative \gls{KL} divergence between
$q_{j}(z_j)$ and $q^*_{j}(z_j)$ from Equation\nobreakspace \textup {(\ref {eq:optimal-qj})}. Thus we
maximize the \gls{ELBO} with respect to $q_j$ when we set
$q_{j}(z_j) = q^*_{j}(z_j)$.

\subsection{Practicalities}
\label{sec:practicalities}

Here, we highlight a few things to keep in mind when implementing and
using variational inference in practice.

\begin{figure}[htb]
  \centering
  \includegraphics[width=5in]{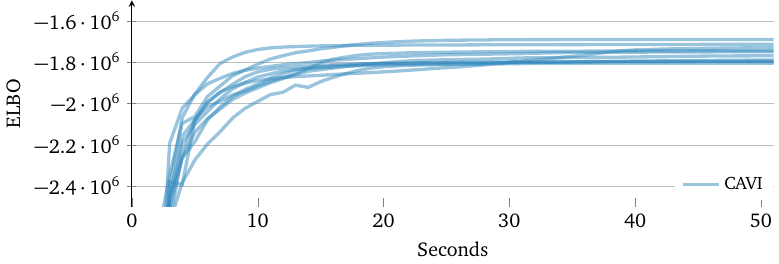}
  \caption{Different initializations may lead \gls{CAVI} to find different local
optima of the \gls{ELBO}.}
  \label{fig:init}
\end{figure}

\parhead{Initialization.}  The \gls{ELBO} is (generally) a non-convex
objective function. \gls{CAVI} only guarantees convergence to a local
optimum, which can be sensitive to initialization. Figure\nobreakspace \ref {fig:init}
shows the \gls{ELBO} trajectory for 10 random initializations using
the Gaussian mixture model.
The means of the variational factors were
randomly initialized by drawing from a factorized Gaussian calibrated
to the empirical mean and variance of the dataset.
(This inference is on images; see
Section\nobreakspace \ref {sec:gaussian-study}.) Each initialization reaches a different
value, indicating the presence of many local optima in the \gls{ELBO}.
In terms of $\gls{KL}(q || p)$, better local optima give variational
densities that are closer to the exact posterior.

This is not always a disadvantage. Some models, such as the mixture of Gaussians
(Section\nobreakspace \ref {sec:mog} and\nobreakspace appendix\nobreakspace \ref {app:cavi}) and mixed-membership model (Appendix\nobreakspace \ref {app:lda}), exhibit
many posterior modes due to label switching: swapping cluster assignment labels
induces many symmetric posterior modes. Representing one of these modes is
sufficient for exploring latent clusters or predicting new observations.

\parhead{Assessing convergence.} Monitoring the \gls{ELBO} in
\gls{CAVI} is simple; we typically assess convergence once the change
in \gls{ELBO} has fallen below some small threshold. However,
computing the \gls{ELBO} of the full dataset may be
undesirable. Instead, we suggest computing the average log predictive
of a small held-out dataset.  Monitoring changes here is a proxy to
monitoring the \gls{ELBO} of the full data.  (Unlike the full
\gls{ELBO}, held-out predictive probability is not guaranteed to
monotonically increase across iterations of \gls{CAVI}.)

\parhead{Numerical stability.} Probabilities are constrained to live
within $ [0,1]$. Precisely manipulating and performing arithmetic of
small numbers requires additional care. When possible, we recommend
working with logarithms of probabilities.  One useful identity is the
``log-sum-exp'' trick,
\begin{align}
  \log \left[ \sum_i \exp( x_i ) \right]
  &=
  \alpha + \log \left[ \sum_i \exp( x_i - \alpha ) \right].
\end{align}
The constant $\alpha$ is typically set to $\max_i x_i$. This provides
numerical stability to common computations in variational inference
procedures.

\section{A complete example: Bayesian mixture of Gaussians}
\label{sec:mog}

As an example, we return to the simple mixture of Gaussians model of
Section\nobreakspace \ref {sec:latent-var-models}. To review, consider $K$ mixture
components and $n$ real-valued data points $x_{1:n}$. The latent
variables are $K$ real-valued mean parameters $\bmu = \mu_{1:K}$ and
$n$ latent-class assignments $\bc = c_{1:n}$. The assignment $c_i$
indicates which latent cluster $x_i$ comes from. In detail, $c_i$ is
an indicator $K$-vector, all zeros except for a one in the position
corresponding to $x_i$'s cluster. There is a fixed hyperparameter
$\sigma^2$, the variance of the normal prior on the $\mu_k$'s. We
assume the observation variance is one and take a uniform prior over
the mixture components.

The joint density of the latent and observed variables is in
Equation\nobreakspace \textup {(\ref {eq:gmm})}.  The variational family is in Equation\nobreakspace \textup {(\ref {eq:gmm-mf-family})}.
Recall that there are two types of variational
parameters---categorical parameters $\varphi_i$ for approximating the
posterior cluster assignment of the $i$th data point and Gaussian
parameters $m_k$ and $s^2_k$ for approximating the posterior of the $k$th
mixture component.

We combine the joint and the mean-field family to form the \gls{ELBO}
for the mixture of Gaussians.  It is a function of the variational
parameters $\bm$, $\bs^2$, and $\bvarphi$,
\begin{align}
\label{eq:mog-elbo}
\begin{split}
\ELBO(\bm, \bs^2, \bvarphi)
  & = \sum_{k=1}^{K}  \EE{\log p(\mu_k); m_k, s^2_k} \\
  & \quad + \sum_{i=1}^{n} \bigl( \EE{\log p(c_i); \varphi_i}
    + \EE{\log p(x_i \g c_i, \bmu); \varphi_i, \bm, \bs^2} \bigr)  \\
  & \quad - \sum_{i=1}^{n} \EE{\log q(c_i; \varphi_i)} - \sum_{k=1}^{K}
      \EE{\log q(\mu_k; m_k, s^2_k)}.
\end{split}
\end{align}
In each term, we have made explicit the dependence on the variational
parameters.  Each expectation can be computed in closed form.

The \gls{CAVI} algorithm updates each variational parameter in turn.
We first derive the update for the variational cluster assignment
factor; we then derive the update for the variational mixture
component factor.

\subsection{The variational density of the mixture assignments}
We first derive the variational update for the cluster assignment
$c_i$.  Using Equation\nobreakspace \textup {(\ref {eq:optimal-qj})},
\begin{align}
  q^*(c_i; \varphi_i) \propto \exp\left\{
  \log p(c_i) + \EE{\log p(x_i \g
  c_i, \bmu); \bm, \bs^2}
  \right\}.
  \label{eq:gmm-z-update}
\end{align}
The terms in the exponent are the components of the joint density
that depend on $c_i$.  The expectation in the second term is over the
mixture components $\bmu$.

The first term of Equation\nobreakspace \textup {(\ref {eq:gmm-z-update})} is the log prior of $c_i$. It
is the same for all possible values of $c_i$,
$\log p(c_i) = - \log K$. The second term is the expected log of the
$c_i$th Gaussian density. Recalling that $c_i$ is an indicator vector,
we can write
$$p(x_i \g c_i, \bmu) = \prod_{k=1}^K p(x_i \g \mu_k)^{c_{ik}}.$$ We
use this to compute the expected log probability,
\begin{align}
  \EE{\log p(x_i \g c_i, \bmu)}
  &= \sum_k c_{ik} \EE{\log p(x_i \g \mu_k); m_k, s^2_k} \\
  &= \sum_k c_{ik} \EE{-(x_i - \mu_k)^2 / 2;  m_k, s^2_k} +
    \const. \\
  &= \sum_k c_{ik} \left(\EE{\mu_k; m_k, s_k^2} x_i - \EE{\mu_k^2;
    m_k, s_k^2} / 2
    \right) + \const. \label{eq:gmm-z-update2}
\end{align}
In each line we remove terms that are constant with respect to $c_i$.
This calculation requires $\EE{\mu_k}$ and $\EE{\mu_k^2}$ for each
mixture component, both computable from the variational Gaussian on
the $k$th mixture component.

Thus the variational update for the $i$th cluster assignment is
\begin{align}
\varphi_{ik} \propto \exp\left\{\EE{\mu_k; m_k,
  s_k^2} x_i - \EE{\mu_k^2; m_k, s_k^2} / 2\right\}.
\end{align}
Notice it is only a function of the variational parameters for the
mixture components.

\subsection{The variational density of the mixture-component means}
\label{sec:vardist-mixcomp-means}

We turn to the variational density
$q(\mu_k; m_k, s_k^2)$ of the $k$th mixture
component. Again we use Equation\nobreakspace \textup {(\ref {eq:optimal-qj})} and write down the joint
density up to a normalizing constant,
\begin{align}
  q(\mu_k) \propto \exp
  \left\{
    \log p(\mu_k) + \textstyle \sum_{i=1}^{n} \EE{\log p(x_i \g c_i,
  \bmu); \varphi_i, \bm_{-k}, \bs^2_{-k}}
  \right\}. \label{eq:cavi-muk}
\end{align}
We now calculate the unnormalized log of this coordinate-optimal
$q(\mu_k)$. Recall $\varphi_{ik}$ is the probability that the $i$th
observation comes from the $k$th cluster. Because $c_i$ is an
indicator vector, we see that $\varphi_{ik} =
\EE{c_{ik}; \varphi_i}$. Now
\begin{align}
  \log q(\mu_k)
  &= \log p(\mu_k) + \textstyle \sum_i \EE{\log p(x_i \g c_i, \bmu);
    \varphi_i, \bm_{-k}, \bs^2_{-k}} + \const. \\
  &= \log p(\mu_k) + \textstyle \sum_i \EE{c_{ik} \log p(x_i \g \mu_k);
    \varphi_i} + \const. \\
  &= - \mu_k^2 / 2 \sigma^2 +
    \textstyle \sum_i \EE{c_{ik}; \varphi_i}
    \log p(x_i \g \mu_{k}) + \const. \\
  &= - \mu_k^2 / 2 \sigma^2 +
    \textstyle \sum_i \varphi_{ik} \left(-(x_i - \mu_k)^2 /
    2 \right) + \const. \\
  &= - \mu_k^2 / 2 \sigma^2 +
    \textstyle \sum_i \varphi_{ik} x_i \mu_k - \varphi_{ik}
    \mu_k^2 /2 + \const. \\
  &= \left(\textstyle \sum_i \varphi_{ik} x_i \right) \mu_k -
    \left(1/2 \sigma^2 + \textstyle \sum_i \varphi_{ik} /
    2\right) \mu_k^2 + \const.
\end{align}
This calculation reveals that the coordinate-optimal variational
density of $\mu_k$ is an exponential family with sufficient
statistics $\{ \mu_k, \mu_k^2 \}$ and natural parameters
$\{ \sum_{i=1}^{n} \varphi_{ik} x_i, -1/2 \sigma^2 - \textstyle
\sum_{i=1}^{n} \varphi_{ik} / 2 \}$,
i.e., a Gaussian.  Expressed in terms of the variational mean and
variance, the updates for $q(\mu_k)$ are
\begin{align}
  m_k
  &= \frac{\sum_i \varphi_{ik} x_i}{1/\sigma^2 + \sum_i \varphi_{ik}},
  \qquad
  s^2_k
  = \frac{1}{1/\sigma^2 + \sum_i \varphi_{ik}}.
  \label{eq:gmm-mu-update}
\end{align}

These updates relate closely to the complete conditional density
of the $k$th component in the mixture model. The complete conditional
is a posterior Gaussian given the data assigned to the $k$th
component. The variational update is a weighted complete conditional,
where each data point is weighted by its variational probability of
being assigned to component $k$.

\begin{algorithm}[t]
\setstretch{1.25}

\KwIn{Data $x_{1:n}$, number of components $K$, prior variance of
  component means $\sigma^2$ }

\KwOut{Variational densities $q(\mu_k; m_k, s^2_k)$ (Gaussian)
 and $q(c_i; \varphi_i)$ ($K$-categorical)}

\textbf{Initialize:} Variational parameters $\bm = m_{1:K}$,
$\bs^2 = s^2_{1:K}$, and $\bvarphi = \varphi_{1:n}$\\

\While{the \gls{ELBO} has not converged} {
  \For{$i \in \{1, \ldots, n\}$}
  {
    Set $\varphi_{ik} \propto \exp\{\EE{\mu_k; m_k,
      s^2_k} x_i - \EE{\mu_k^2; m_k, s^2_k} / 2\}$\\
  }
  \For{$k \in \{1, \ldots, K\}$}
  {
    \begin{flalign*}
      \mathrm{Set \,} m_k & \longleftarrow \frac{\sum_i
        \varphi_{ik} x_i}{1/\sigma^2 + \sum_i \varphi_{ik}} & \\
      \mathrm{Set \,} s^2_k & \longleftarrow \displaystyle
      \frac{1}{1/\sigma^2 + \sum_i \varphi_{ik}} &
    \end{flalign*}
  }
  Compute $\ELBO(\bm, \bs^2, \bvarphi)$
}
\Return{$q(\bm, \bs^2, \bvarphi)$}
\caption{\gls{CAVI} for a Gaussian mixture model}
\label{alg:gmm-cavi}
\end{algorithm}

\subsection{CAVI for the mixture of Gaussians}

Algorithm\nobreakspace \ref {alg:gmm-cavi} presents coordinate-ascent variational inference
for the Bayesian mixture of Gaussians.  It combines the variational
updates in Equation\nobreakspace \textup {(\ref {eq:gmm-z-update})} and Equation\nobreakspace \textup {(\ref {eq:gmm-mu-update})}.  The
algorithm requires computing the \gls{ELBO} of Equation\nobreakspace \textup {(\ref {eq:mog-elbo})}.  We
use the \gls{ELBO} to track the progress of the algorithm and assess
when it has converged.

Once we have a fitted variational density, we can use it as we
would use the posterior. For example, we can obtain a posterior
decomposition of the data. We assign points to their most likely
mixture assignment $\hat{c}_i = \argmax_{k} \varphi_{ik}$ and estimate
cluster means with their variational means $m_k$.

We can also use the fitted variational density to approximate the
predictive density of new data.  This approximate predictive is a
mixture of Gaussians,
\begin{align}
  p(x_{\textrm{new}} \g x_{1:n}) \approx
  \frac{1}{K} \sum_{k=1}^{K} p(x_{\textrm{new}} \g m_k),
\end{align}
where $p(x_{\textrm{new}} \g m_k)$ is a Gaussian with
mean $m_k$ and unit variance.

\subsection{Empirical study}
\label{sec:gaussian-study}

We present two analyses to demonstrate the mixture of Gaussians
algorithm in action.  The first is a simulation study; the second is
an analysis of a data set of natural images.

\begin{figure}[p]
  \centering
  \includegraphics[width=5in]{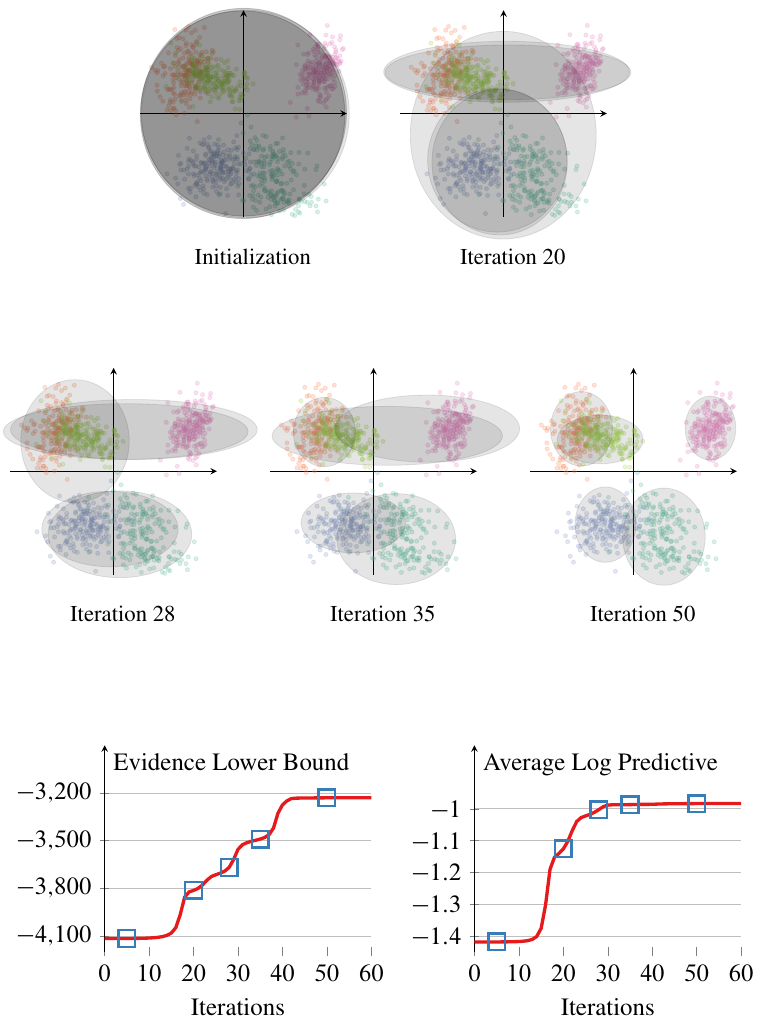}
  \caption{A simulation study of a two dimensional Gaussian mixture model.
  The ellipses are $2\sigma$ contours of the variational approximating factors.}
  \label{fig:gmm_2d}
\end{figure}

\parhead{Simulation study.} Consider two-dimensional real-valued data
$\bx$. We simulate $K=5$ Gaussians with random means, covariances, and
mixture assignments. Figure\nobreakspace \ref {fig:gmm_2d} shows the data; each point is
colored according to its true cluster.  Figure\nobreakspace \ref {fig:gmm_2d} also
illustrates the initial variational density of the mixture
components---each is a Gaussian, nearly centered, and with a wide
variance; the subpanels plot the variational density of the
components as the \gls{CAVI} algorithm progresses.

The progression of the \gls{ELBO} tells a story.  We highlight key
points where the \gls{ELBO} develops ``elbows'', phases of the
maximization where the variational approximation changes its
shape. These ``elbows'' arise because the \gls{ELBO} is not a convex
function in terms of the variational parameters; \gls{CAVI}
iteratively reaches better plateaus.

Finally, we plot the logarithm of the Bayesian predictive density
as approximated by the variational density. Here we report the
average across held-out data. Note this plot is smoother than
the \gls{ELBO}.

\begin{figure}[htb]
  \centering
  \includegraphics[width=5in]{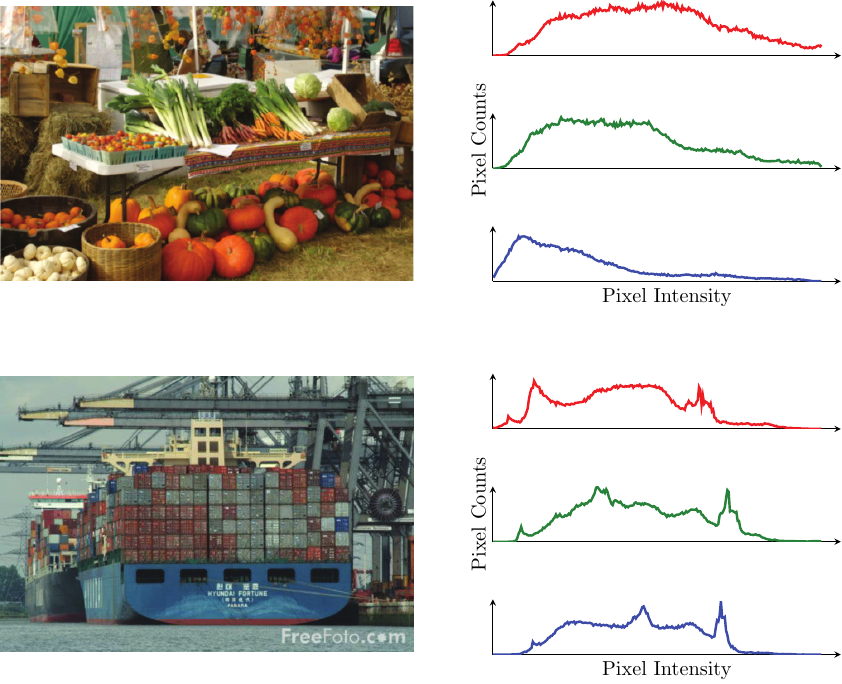}
  \caption{Red, green, and blue channel image histograms for two images from
  the image\textsc{clef} dataset. The top image lacks blue hues, which is
  reflected in its blue channel histogram. The bottom image has a few
  dominant shades of blue and green, as seen in the peaks of
  its histogram.}
  \label{fig:histo}
\end{figure}

\parhead{Image analysis.}  We now turn to an experimental
study. Consider the task of grouping images according to their color
profiles. One approach is to compute the color histogram of the
images. Figure\nobreakspace \ref {fig:histo} shows the red, green, and blue channel
histograms of two images from the image\textsc{clef} data
\citep{villegas13_CLEF}. Each histogram is a vector of length 192;
concatenating the three color histograms gives a 576-dimensional
representation of each image, regardless of its original size in
pixel-space.

\begin{figure}[htb]
\centering
  \begin{subfigure}[b]{1in}
    \centering
    \includegraphics[width=1in]{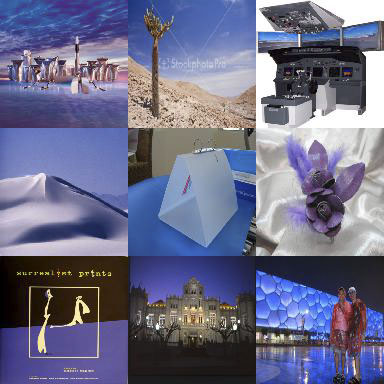}
    \caption{Purple}
    \label{sub:purple}
  \end{subfigure}
  \hspace*{0.1in}
  \begin{subfigure}[b]{1in}
    \centering
    \includegraphics[width=1in]{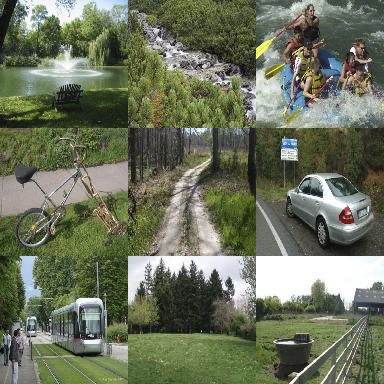}
    \caption{Green \& White}
    \label{sub:green_and_white}
  \end{subfigure}
  \hspace*{0.1in}
  \begin{subfigure}[b]{1in}
    \centering
    \includegraphics[width=1in]{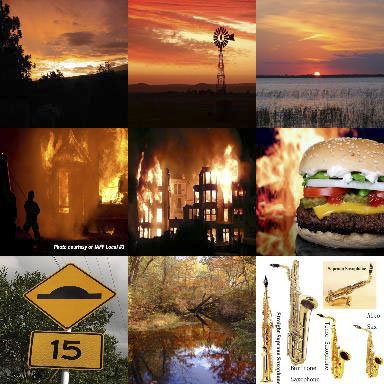}
    \caption{Orange}
    \label{sub:orange}
  \end{subfigure}
  \hspace*{0.1in}
  \begin{subfigure}[b]{1in}
    \centering
    \includegraphics[width=1in]{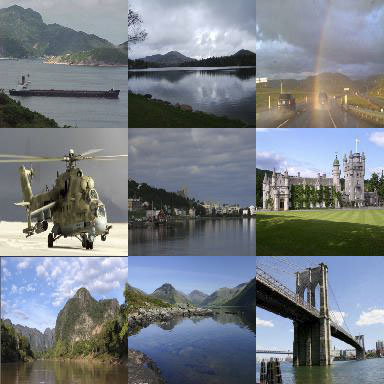}
    \caption{Grayish Blue}
    \label{sub:grayish_blue}
  \end{subfigure}
  \caption{Example clusters from the Gaussian mixture model. We
  assign each image to its most likely mixture cluster. The subfigures show nine
  randomly sampled images from four clusters; their namings are subjective.
  }
  \label{fig:clusters}
\end{figure}

\begin{figure}[htb]
  \centering
  \includegraphics[width=5in]{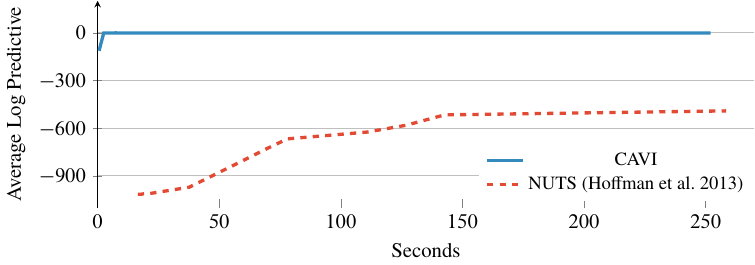}
  \caption{Comparison of \gls{CAVI} to a Hamiltonian Monte Carlo-based sampling
  technique. \gls{CAVI} fits a Gaussian mixture model to ten
  thousand images in less than a minute.}
  \label{fig:image_logpred}
\end{figure}

We use \gls{CAVI} to fit a Gaussian mixture model with thirty clusters
to image
histograms. We randomly select two sets of ten thousand images from
the image\textsc{clef} collection to serve as training and testing
datasets.  Figure\nobreakspace \ref {fig:clusters} shows similarly colored images assigned
to four randomly chosen clusters.  Figure\nobreakspace \ref {fig:image_logpred} shows the
average log predictive accuracy of the testing set as a function of
time. We compare \gls{CAVI} to an implementation in Stan
\citep{stan-manual:2015}, which uses a Hamiltonian Monte Carlo-based
sampler \citep{Hoffman-Gelman:2011}. (Details are in Appendix\nobreakspace \ref {app:cavi}.)
\gls{CAVI} is orders of magnitude faster than this sampling
algorithm.\footnote{This is not a definitive comparison between
  variational inference and \gls{MCMC}.  Other samplers, such as a collapsed
  Gibbs sampler, may perform better than Hamiltonian Monte Carlo sampling.}

\section{Variational inference with exponential
  families} \label{sec:ef}

We described mean-field variational inference and derived \gls{CAVI},
a general coordinate-ascent algorithm for optimizing the \gls{ELBO}.
We demonstrated this approach on a simple mixture of Gaussians, where
each coordinate update was available in closed form.

The mixture of Gaussians is one member of the important class of
models where each complete conditional is in the exponential family.
This includes a number of widely used models, such as Bayesian
mixtures of exponential families, factorial mixture models, matrix
factorization models, certain hierarchical regression models (e.g.,
linear regression, probit regression), stochastic
blockmodels of networks, hierarchical mixtures of experts, and a
variety of mixed-membership models (which we will discuss below).

Working in this family simplifies variational inference: it is easier
to derive the corresponding \gls{CAVI} algorithm, and it enables
variational inference to scale up to massive data.  In
Section\nobreakspace \ref {sec:exp-fam}, we develop the general case.  In
Section\nobreakspace \ref {sec:cond-conj}, we discuss conditionally conjugate models, i.e.,
the common Bayesian application where some latent variables are
``local'' to a data point and others, usually identified with
parameters, are ``global'' to the entire data set.  Finally, in
Section\nobreakspace \ref {sec:svi}, we describe stochastic variational
inference~\citep{Hoffman:2013}, a stochastic optimization algorithm
that scales up variational inference in this setting.

\subsection{Complete conditionals in the exponential family}
\label{sec:exp-fam}

Consider the generic model $p(\bz, \bx)$ of
Section\nobreakspace \ref {sec:latent-var-models} and suppose each complete conditional is
in the exponential family:
\begin{align}
  \label{eq:ef-conditional}
  p(z_j \g \bz_{-j}, \bx) =
  h(z_j) \exp \{
  \eta_j(\bz_{-j}, \bx)^\top z_j - a(\eta_j(\bz_{-j},
  \bx)) \},
\end{align}
where $z_j$ is its own sufficient statistic, $h(\cdot)$ is a base
measure, and $a(\cdot)$ is the log normalizer~\citep{Brown:1986}.
Because this is a conditional density, the parameter
$\eta_j(\bz_{-j}, \bx)$ is a function of the conditioning set.

Consider mean-field variational inference for this class of models,
where we fit $q(\bz) = \prod_{j} q_j(z_j)$.  The exponential family
assumption simplifies the coordinate update of Equation\nobreakspace \textup {(\ref {eq:qstar})},
\begin{align}
  q(z_j)
  &\propto \exp\left\{\EE{\log p(z_j \g \bz_{-j}, \bx)}\right\} \\
  &= \exp
      \left\{
      \log h(z_j) + \EE{\eta_j(\bz_{-j}, \bx)}^\top z_j -
      \EE{a(\eta_j(\bz_{-j}, \bx))}
      \right\} \\
  &\propto h(z_j) \exp\left\{ \EE{\eta_j(\bz_{-j}, \bx)}^\top z_j \right\}.
\end{align}
This update reveals the parametric form of the optimal variational
factors.  Each one is in the same exponential family as its
corresponding complete conditional.  Its parameter has the same
dimension and it has the same base measure $h(\cdot)$ and log
normalizer $a(\cdot)$.

Having established their parametric forms, let $\nu_j$ denote the
variational parameter for the $j$th variational factor.  When we
update each factor, we set its parameter equal to the expected
parameter of the complete conditional,
\begin{align}
  \label{eq:exp-fam-cavi}
  \nu_j = \EE{\eta_j(\bz_{-j}, \bx)}.
\end{align}
This expression facilitates deriving \gls{CAVI} algorithms for many
complex models.

\subsection{Conditional conjugacy and Bayesian models}
\label{sec:cond-conj}

One important special case of exponential family models are
\textit{conditionally conjugate models} with local and global
variables. Models like this come up frequently in Bayesian statistics
and statistical machine learning, where the global variables are the
``parameters'' and the local variables are per-data-point latent
variables.

\parhead{Conditionally conjugate models.}  Let $\beta$ be a vector of
\textit{global latent variables}, which potentially govern any of the
data. Let $\bz$ be a vector of \textit{local latent variables}, whose
$i$th component only governs data in the $i$th ``context.'' The joint
density is
\begin{align}
  \label{eq:lg-joint}
  p(\beta, \bz, \bx) = p(\beta) \prod_{i=1}^{n} p(z_i, x_i \g \beta).
\end{align}
The mixture of Gaussians of Section\nobreakspace \ref {sec:mog} is an example. The global
variables are the mixture components; the $i$th local variable is the
cluster assignment for data point $x_i$.

We will assume that the modeling terms of~Equation\nobreakspace \textup {(\ref {eq:lg-joint})} are
chosen to ensure each complete conditional is in the exponential
family. In detail, we first assume the joint density of each
$(x_i, z_i)$ pair, conditional on $\beta$, has an exponential family
form,
\begin{align}
  \label{eq:zi-xi-given-beta}
  p(z_i, x_i \g \beta) = h(z_i, x_i) \exp\{\beta^\top t(z_i, x_i) - a(\beta)\},
\end{align}
where $t(\cdot, \cdot)$ is the sufficient statistic.

Next, we take the prior on the global variables to be the
corresponding conjugate prior~\citep{diaconis1979conjugate,Bernardo:1994},
\begin{align}
  p(\beta) = h(\beta) \exp\{\alpha^\top [\beta, -a(\beta)]- a(\alpha)\}.
\end{align}
This prior has natural (hyper)parameter
$\alpha = [\alpha_1, \alpha_2]^\top$, a column vector, and sufficient statistics
that concatenate the global variable and its log normalizer in the
density of the local variables.

With the conjugate prior, the complete conditional of the global
variables is in the same family.  Its natural parameter is
\begin{align}
  \hat{\alpha}
  &=
    \begin{bmatrix}
      \alpha_1 + \textstyle \sum_{i=1}^{n} t(z_i,x_i), \:
      \alpha_2 + n
    \end{bmatrix}^\top.
  \label{eq:hat-alpha}
\end{align}

Turn now to the complete conditional of the local variable
$z_i$. Given $\beta$ and $x_i$, the local variable $z_i$ is
conditionally independent of the other local variables $\bz_{-i}$ and
other data $\bx_{-i}$.  This follows from the form of the joint
density in Equation\nobreakspace \textup {(\ref {eq:lg-joint})}.  Thus
\begin{align}
  \label{eq:local-cc}
  p(z_i \g x_i, \beta, \bz_{-i}, \bx_{-i}) = p(z_i \g x_i, \beta).
\end{align}
We further assume that this density is in an exponential family,
\begin{align}
  p(z_i \g x_i, \beta) =
  h(z_i) \exp\{\eta(\beta, x_i)^\top z_i - a(\eta(\beta, x_i))\}.
\end{align}
This is a property of the local likelihood term $p(z_i, x_i \g \beta)$
from~Equation\nobreakspace \textup {(\ref {eq:zi-xi-given-beta})}.  For example, in the mixture of
Gaussians, the complete conditional of the local variable is a
categorical.

\parhead{Variational inference in conditionally conjugate models.}  We
now describe \gls{CAVI} for this general class of models. Write
$q(\beta \g \lambda)$ for the variational posterior approximation on
$\beta$; we call $\lambda$ the ``global variational parameter''.  It
indexes the same exponential family density as the prior.
Similarly, let the variational posterior $q(z_i \g \varphi_i)$ on each
local variable $z_i$ be governed by a ``local variational parameter''
$\varphi_{i}$.  It indexes the same exponential family density as
the local complete conditional.  \gls{CAVI} iterates between updating
each local variational parameter and updating the global variational
parameter.

The local variational update is
\begin{align}
  \label{eq:local-update}
  \varphi_i = \E_{\lambda}\left[\eta(\beta, x_i)\right].
\end{align}
This is an application of Equation\nobreakspace \textup {(\ref {eq:exp-fam-cavi})}, where we take the
expectation of the natural parameter of the complete conditional in
Equation\nobreakspace \textup {(\ref {eq:local-cc})}.

The global variational update applies the same technique.  It is
\begin{align}
  \lambda
  &=
  \begin{bmatrix}
    \alpha_1 + \textstyle \sum_{i=1}^{n} \E_{\varphi_i}\left[t(z_i,
    x_i)\right], \:
    \alpha_2 + n
  \end{bmatrix}^\top.
  \label{eq:global-update}
\end{align}
Here we take the expectation of the natural parameter in
Equation\nobreakspace \textup {(\ref {eq:hat-alpha})}.

\gls{CAVI} optimizes the \gls{ELBO} by iterating between local updates
of each local parameter and global updates of the global parameters.
To assess convergence we can compute the \gls{ELBO} at each iteration
(or at some lag), up to a constant that does not depend on the
variational parameters,
\begin{align}
  \ELBO =
  \left(\alpha_1 + \textstyle \sum_{i=1}^{n} \E_{\varphi_i}\left[
  t(z_i, x_i) \right]\right)^\top \E_{\lambda}\left[\beta\right] - (\alpha_2 + n) \E_{\lambda}\left[ a(\beta) \right] -
  \EE{\log q(\beta, \bz)}.
  \label{eq:cond-conj-elbo}
\end{align}
This is the \gls{ELBO} in Equation\nobreakspace \textup {(\ref {eq:elbo})} applied to the joint in
Equation\nobreakspace \textup {(\ref {eq:lg-joint})} and the corresponding mean-field variational
density; we have omitted terms that do not depend on the
variational parameters.  The last term is
\begin{align}
  \EE{\log q(\beta, \bz)}
  =
    \lambda^\top \E_{\lambda}\left[ t(\beta) \right] - a(\lambda) + \sum_
    {i=1}^{n}
    \varphi_i^\top \E_{\varphi_i} \left[ z_i \right] - a(\varphi_i).
\end{align}
\gls{CAVI} for the mixture of Gaussians model (Algorithm\nobreakspace \ref {alg:gmm-cavi}) is
an instance of this method. Appendix\nobreakspace \ref {app:lda} presents another example of
\gls{CAVI} for \gls{LDA}, a probabilistic topic model.

\subsection{Stochastic variational inference}
\label{sec:svi}

Modern applications of probability models often require analyzing
massive data.  However, most posterior inference algorithms do not
easily scale. \gls{CAVI} is no exception, particularly in the
conditionally conjugate setting of Section\nobreakspace \ref {sec:cond-conj}.  The reason
is that the coordinate ascent structure of the algorithm requires
iterating through the entire data set at each iteration.  As the data
set size grows, each iteration becomes more computationally expensive.

An alternative to coordinate ascent is gradient-based optimization,
which climbs the \gls{ELBO} by computing and following its gradient at
each iteration. This perspective is the key to scaling up variational
inference using \gls{SVI}~\citep{Hoffman:2013}, a method that combines
natural gradients~\citep{Amari:1998} and stochastic
optimization~\citep{Robbins:1951}.

\gls{SVI} focuses on optimizing the global variational parameters
$\lambda$ of a conditionally conjugate model. The flow of computation
is simple. The algorithm maintains a current estimate of the global
variational parameters. It repeatedly (a) subsamples a data point from
the full data set; (b) uses the current global parameters to compute
the optimal local parameters for the subsampled data point; and (c)
adjusts the current global parameters in an appropriate way. \gls{SVI}
is detailed in Algorithm\nobreakspace \ref {alg:svi}.  We now show why it is a valid
algorithm for optimizing the \gls{ELBO}.

\parhead{The natural gradient of the \gls{ELBO}.}  In gradient-based
optimization, the \textit{natural gradient} accounts for the geometric
structure of probability parameters~\citep{Amari:1982,Amari:1998}.
Specifically, natural gradients warp the parameter space in a sensible
way, so that moving the same distance in different directions amounts
to equal change in symmetrized \gls{KL} divergence.  The usual
Euclidean gradient does not enjoy this property.

In exponential families, we find the natural gradient with respect to
the parameter by premultiplying the usual gradient by the inverse
covariance of the sufficient statistic, $a''(\lambda)^{-1}$.  This is
the inverse Riemannian metric and the inverse Fisher information
matrix~\citep{Amari:1982}.

Conditionally conjugate models enjoy simple natural gradients of the
\gls{ELBO}.  We focus on gradients with respect to the global
parameter $\lambda$.  \citet{Hoffman:2013} derive the Euclidean
gradient of the \gls{ELBO},
\begin{align}
  \nabla_{\lambda} \ELBO = a''(\lambda)(\EEE{\varphi}{\hat{\alpha}} - \lambda),
\end{align}
where $\EEE{\varphi}{\hat{\alpha}}$ is in Equation\nobreakspace \textup {(\ref {eq:global-update})}.
Premultiplying by the inverse Fisher information gives the natural
gradient $g(\lambda)$,
\begin{align}
  \label{eq:nat-grad}
  g(\lambda) = \EEE{\varphi}{\hat{\alpha}} - \lambda.
\end{align}
It is the difference between the coordinate updates
$\EEE{\varphi}{\hat{\alpha}}$ and the variational parameters $\lambda$
at which we are evaluating the gradient.  In addition to enjoying good
theoretical properties, the natural gradient is easier to calculate
than the Euclidean gradient.  For more on natural gradients and
variational inference see~\citet{Sato:2001} and \citet{Honkela:2008}.

We can use this natural gradient in a gradient-based optimization
algorithm.  At each iteration, we update the global parameters,
\begin{align}
  \lambda_t = \lambda_{t-1} + \epsilon_t g(\lambda_{t-1}),
\end{align}
where $\epsilon_t$ is a step size.

Substituting Equation\nobreakspace \textup {(\ref {eq:nat-grad})} into the second term reveals a special
structure,
\begin{align}
  \lambda_t = (1 - \epsilon_t) \lambda_{t-1} + \epsilon_t
  \EEE{\varphi}{\hat{\alpha}}.
\end{align}
Notice this does not require additional types of calculations other
than those for coordinate ascent updates.  At each iteration, we first
compute the coordinate update. We then adjust the current estimate to
be a weighted combination of the update and the current variational
parameter.

Though easy to compute, using the natural gradient has the same cost
as the coordinate update in Equation\nobreakspace \textup {(\ref {eq:global-update})}; it requires summing over
the entire data set and computing the optimal local variational
parameters for each data point.  With massive data, this is prohibitively expensive.

\parhead{Stochastic optimization of the \gls{ELBO}.} Stochastic
variational inference solves this problem by using the natural
gradient in a stochastic optimization algorithm. Stochastic
optimization algorithms follow noisy but cheap-to-compute gradients to
reach the optimum of an objective function.  (In the case of the
\gls{ELBO}, stochastic optimization will reach a local optimum.) In
their seminal paper, \citet{Robbins:1951} proved results implying that
optimization algorithms can successfully use noisy, unbiased
gradients, as long as the step size sequence satisfies certain
conditions.  This idea has
blossomed~\citep{Spall:2003,Kushner:1997}. Stochastic optimization has
enabled modern machine learning to scale to massive
data~\citep{Bottou:2004}.

Our aim is to construct a cheaply computed, noisy, unbiased natural
gradient. We expand the natural gradient in Equation\nobreakspace \textup {(\ref {eq:nat-grad})} using
Equation\nobreakspace \textup {(\ref {eq:hat-alpha})}
\begin{align}
  g(\lambda) = \alpha + \left[\textstyle \sum_{i=1}^{n}
  \EEE{\varphi^*_i}{t(z_i, x_i)}, \: n\right]^\top -
  \lambda,
\end{align}
where $\varphi^*_i$ indicates that we consider the optimized local
variational parameters (at fixed global parameters $\lambda$) in
Equation\nobreakspace \textup {(\ref {eq:local-update})}.  We construct a noisy natural gradient by
sampling an index from the data and then rescaling the second term,
\begin{align}
  t  &\sim \textrm{Unif}(1, \ldots, n) \\
  \label{eq:noisy-nat-grad}
  \hat{g}(\lambda)
     &=
       \alpha + n \left[\EEE{\varphi^*_t}{t(z_t, x_t)}, \: 1\right]^\top -
       \lambda.
\end{align}
The noisy natural gradient $\hat{g}(\lambda)$ is unbiased:
$\EEE{t}{\hat{g}(\lambda)} = g(\lambda)$.  And it is cheap to
compute---it only involves a single sampled data point and only one
set of optimized local parameters.  (This immediately extends to
minibatches, where we sample $B$ data points and rescale
appropriately.)  Again, the noisy gradient only requires calculations
from the coordinate ascent algorithm.  The first two terms of
Equation\nobreakspace \textup {(\ref {eq:noisy-nat-grad})} are equivalent to the coordinate update in a
model with $n$ replicates of the sampled data point.

Finally, we set the step size sequence.  It must follow the conditions
of~\citet{Robbins:1951},
\begin{align}
  \sum_{t} \epsilon_t = \infty \quad ; \quad \sum_{t} \epsilon_t^2 <
  \infty.
\end{align}
Many sequences will satisfy these conditions, for example
$\epsilon_t = t^{-\kappa}$ for $\kappa \in (0.5, 1]$.  The full
\gls{SVI} algorithm is in Algorithm\nobreakspace \ref {alg:svi}.

We emphasize that \gls{SVI} requires no new derivation beyond what is
needed for \gls{CAVI}. Any implementation of \gls{CAVI} can be
immediately scaled up to a stochastic algorithm.

\parhead{Probabilistic topic models.} We demonstrate \gls{SVI} with a
probabilistic topic model.  Probabilistic topic models are
mixed-membership models of text, used to uncover the latent ``topics''
that run through a collection of documents. Topic models have become a
popular technique for exploratory data analysis of large
collections~\citep{Blei:2012}.

In detail, each latent topic is a distribution over terms in a
vocabulary and each document is a collection of words that comes from
a mixture of the topics.  The topics are shared across the collection,
but each document mixes them with different proportions. (This is the
hallmark of a mixed-membership model.)  Thus topic modeling casts
topic discovery as a posterior inference problem. Posterior estimates
of the topics and topic proportions can be used to summarize,
visualize, explore, and form predictions about the documents.

\begin{figure}[t]
  \centering
    \includegraphics[width=0.75\textwidth]{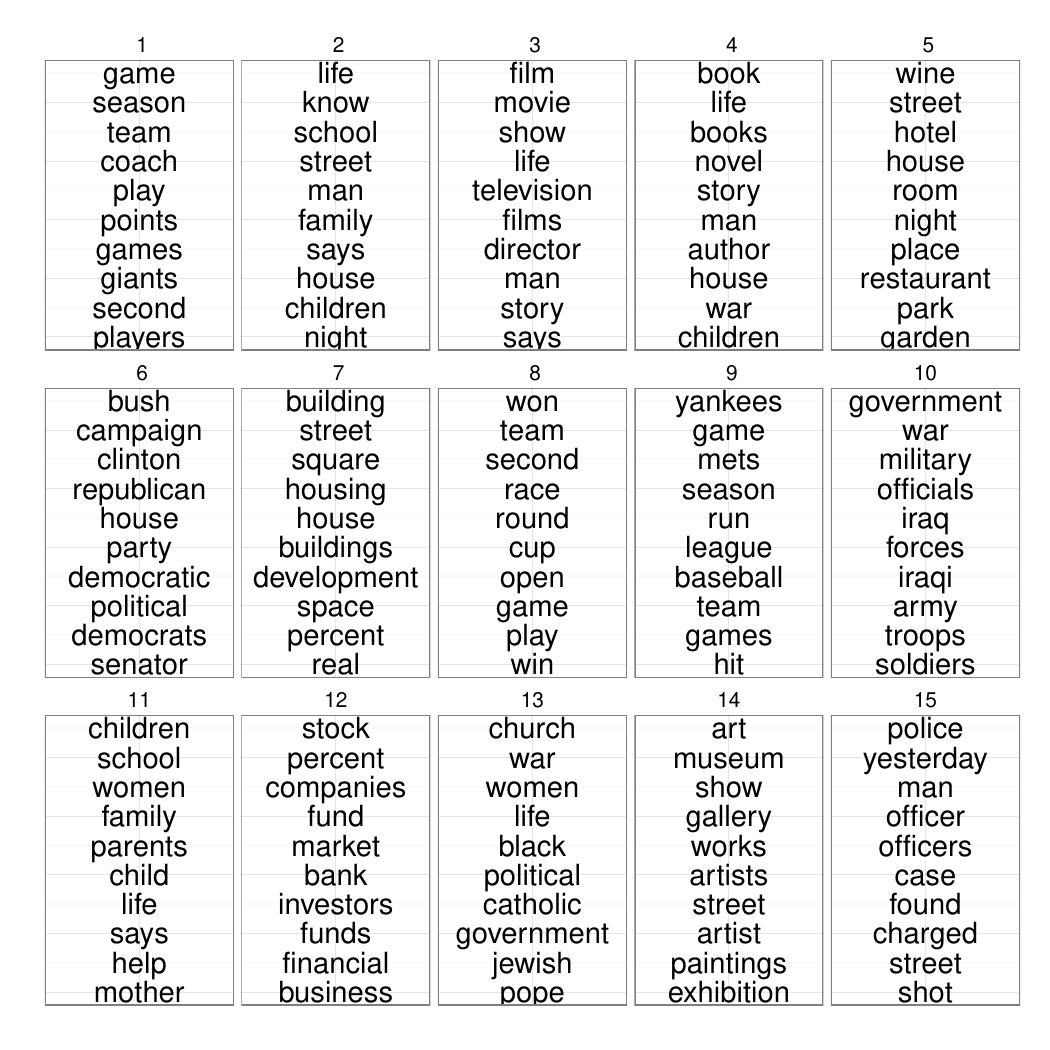}
  \caption{\label{fig:nyt} Topics found in a corpus of 1.8M
    articles from the New York Times. Reproduced with permission
    from~\citet{Hoffman:2013}.}
\end{figure}

One motivation for topic modeling is to get a handle on massive
collections of documents.  Early inference algorithms were based on
coordinate ascent variational inference~\citep{Blei:2003b} and
analyzed collections in the thousands or tens of thousands of
documents.
(Appendix\nobreakspace \ref {app:lda} presents this algorithm).
With \gls{SVI}, topic models scale up to millions of
documents; the details of the algorithm are in~\citet{Hoffman:2013}.
Figure\nobreakspace \ref {fig:nyt} illustrates topics inferred using the latent Dirichlet allocation
model~\citep{Blei:2003b}
from 1.8M articles from the
\textit{New York Times}.  This analysis would not have been possible
without \gls{SVI}.

\begin{algorithm}[t]
\setstretch{1.25}
\KwIn{Model $p(\bx,\bz)$, data $\bx$, and step size sequence $\epsilon_t$}
\KwOut{Global variational densities $q_{\lambda}(\beta)$}
\textbf{Initialize:} Variational parameters $\lambda_0$\\
\While{\textrm{TRUE}}{
  Choose a data point uniformly at random, $t \sim \textrm{Unif}(1,
  \ldots, n)$ \\
  Optimize its local variational parameters $\varphi^*_t =
  \E_{\lambda}\left[\eta(\beta, x_t)\right]$ \\
  Compute the coordinate update as though $x_t$ was repeated $n$ times,
  \begin{equation*}
    \hat{\lambda} = \alpha + n \EEE{\varphi^*_t}{f(z_t, x_t)}
  \end{equation*} \\
  Update the global variational parameter, $\lambda_t = (1 -
  \epsilon_t) \lambda_t + \epsilon_t \hat{\lambda}_t$ \\
}
\Return{$\lambda$}
\caption{\gls{SVI} for conditionally conjugate models}
\label{alg:svi}
\end{algorithm}

\section{Discussion}
\label{sec:discussion}

We described variational inference, a method that uses optimization to
make probabilistic computations.  The goal is to approximate the
conditional density of latent variables $\bz$ given observed
variables $\bx$, $p(\bz \g \bx)$.  The idea is to posit a family of
densities $\cQ$ and then to find the member $q^*(\cdot)$ that is
closest in \gls{KL} divergence to the conditional of interest.
Minimizing the \gls{KL} divergence is the optimization problem, and
its complexity is governed by the complexity of the approximating
family.

We then described the mean-field family, i.e., the family of fully
factorized densities of the latent variables.  Using this family,
variational inference is particularly amenable to coordinate-ascent
optimization, which iteratively optimizes each factor. (This approach
closely connects to the classical Gibbs
sampler~\citep{Geman:1984,Gelfand:1990}.)  We showed how to use
mean-field \gls{VI} to approximate the posterior density of a
Bayesian mixture of Gaussians, discussed the special case of
exponential families and conditional conjugacy, and described the
extension to stochastic variational inference~\citep{Hoffman:2013},
which scales mean-field variational inference to massive data.

\subsection{Applications}

Researchers in many fields have used variational inference to solve
real problems. Here we focus on example applications of mean-field
variational inference and structured variational inference based on
the \gls{KL} divergence.  This discussion is not exhaustive; our intention
is to outline the diversity of applications of variational inference.

\parhead{Computational biology.} \gls{VI} is widely used in
computational biology, where probabilistic models provide important
building blocks for analyzing genetic data.  For example, \gls{VI} has
been used in genome-wide association
studies~\citep{Carbonetto:2012,Logsdon:2010}, regulatory network
analysis~\citep{Sanguinetti:2006}, motif detection~\citep{Xing:2004},
phylogenetic hidden Markov models~\citep{Jojic:2004a}, population
genetics~\citep{Raj:2014}, and gene expression
analysis~\citep{Stegle:2010}.

\parhead{Computer vision and robotics.}  Since its inception,
variational inference has been important to computer vision.  Vision
researchers frequently analyze large and high-dimensional data sets of
images, and fast inference is important to successfully deploy a
vision system.  Some of the earliest examples included inferring
non-linear image manifolds~\citep{Bishop:2000} and finding layers of
images in videos~\citep{Jojic:2001}.  As other examples, variational
inference is important to probabilistic models of
videos~\citep{Chan:2009,Wang:2009c}, image
denoising~\citep{Likas:2004}, tracking~\citep{Vermaak:2003,Yu:2005},
place recognition and mapping for
robotics~\citep{Cummins:2008,Ramos:2012}, and image segmentation with
Bayesian nonparametrics~\citep{Sudderth:2008}.  \citet{Du:2009a} uses
variational inference in a probabilistic model to combine the tasks of
segmentation, clustering, and annotation.

\parhead{Computational neuroscience.} Modern neuroscience research
also requires analyzing very large and high-dimensional data sets,
such as high-frequency time series data or high-resolution functional
magnetic imaging data.  There have been many applications of
variational inference to neuroscience, especially for autoregressive
processes~\citep{Roberts:2002,Penny:2003,Penny:2005,Flandin:2007,Harrison:2010}.
Other applications of variational inference to neuroscience include
hierarchical models of multiple subjects~\citep{Woolrich:2004},
spatial
models~\citep{Sato:2004,Zumer:2007,Kiebel:2008,Wipf:2009a,Lashkari:2012,Nathoo:2014},
brain-computer interfaces~\citep{Sykacek:2004}, and factor
models~\citep{Manning:2014,Gershman:2014}.  There is a software
toolbox that uses variational methods for solving neuroscience and
psychology research problems~\citep{Daunizeau:2014}.

\parhead{Natural language processing and speech recognition.}  In
natural language processing, variational inference has been used for
solving problems such as parsing~\citep{Liang:2007,Liang:2009b},
grammar induction~\citep{Kurihara:2006,Naseem:2010,Cohen:2010a},
models of streaming text~\citep{Yogatama:2014}, topic
modeling~\citep{Blei:2003b}, and hidden Markov models and part-of-speech
tagging~\citep{Wang:2013b}.  In speech recognition, variational
inference has been used to fit complex coupled
hidden Markov models~\citep{Reyes-Gomez:2004} and switching dynamic
systems~\citep{Deng:2004}.

\parhead{Other applications.}  There have been many other applications
of variational inference.  Fields in which it has been used include
marketing~\citep{Braun:2010}, optimal control and reinforcement
learning~\citep{Van-Den-Broek:2008, Furmston:2010}, statistical
network analysis~\citep{Wiggins:2008,Airoldi:2008},
astrophysics~\citep{Regier:2015}, and the social
sciences~\citep{Erosheva:2007,Grimmer:2010a}.  General variational
inference algorithms have been developed for a variety of classes of
models, including shrinkage
models~\citep{Armagan:2011a,Armagan:2011,Neville:2014}, general
time-series
models~\citep{Roberts:2004a,Barber:2006,Archambeau:2007,Archambeau:2007a,Johnson:2014,Foti:2014},
robust models~\citep{Tipping:2005,Wang:2015}, and Gaussian process
models~\citep{Titsias:2010,Damianou:2011,Hensman:2014}.

\subsection{Theory}
\label{sec:theory}

Though researchers have not developed much theory around variational
inference, there are several threads of research about theoretical
guarantees of variational approximations. As we mentioned in the
introduction, one of our purposes for writing this paper is to
catalyze research on the statistical theory around variational
inference.

Below, we summarize a variety of results.  In general, they are all of
the following type: treat \gls{VI} posterior means as point estimates
(or use \textsc{m}-step estimates from variational \gls{EM}) and
confirm that they have the usual frequentist asymptotics. (Sometimes
the research finds that they do not enjoy the same asymptotics.) Each
result revolves around a single model and a single family of
variational approximations.

\citet{you2014variational} study the variational posterior for a
classical Bayesian linear model. They put a normal prior on the
coefficients and an inverse gamma prior on the response variance. They
find that, under standard regularity conditions, the mean-field
variational posterior mean of the parameters are consistent in the
frequentist sense.  \citet{you2014} build on their earlier work with a
spike-and-slab prior on the coefficients and find similar consistency
results.

\citet{Hall:2011a,Hall:2011} examine a simple Poisson mixed-effects
model, one with a single predictor and a random intercept. They use a
Gaussian variational approximation and estimate parameters with
variational \gls{EM}.  They prove consistency of these estimates at
the parametric rate and show asymptotic normality with asymptotically
valid standard errors.

\citet{celisse2012consistency} and \citet{bickel2013asymptotic}
analyze network data using stochastic blockmodels. They show
asymptotic normality of parameter estimates obtained using a
mean-field variational approximation. They highlight the computational
advantages and theoretical guarantees of the variational approach over
maximum likelihood for dense, sparse, and restricted variants of the
stochastic blockmodel.

\citet{westling2015establishing} study the consistency of \gls{VI}
through a connection to M-estimation. They focus on a broader class of
models (with posterior support in real coordinate space) and analyze
an automated \gls{VI} technique that uses a Gaussian variational
approximation \citep{kucukelbir2015automatic}. They derive an
asymptotic covariance matrix estimator of the variational approximation
and show its robustness to model misspecification.

Finally, \citet{Wang:2006} analyze variational approximations to
mixtures of Gaussians. Specifically, they consider Bayesian mixtures
with conjugate priors, the mean-field variational approximation, and
an estimator that is the variational posterior mean. They confirm that
\gls{CAVI} converges to a local optimum, that the \gls{VI} estimator
is consistent, and that the \gls{VI} estimate and \gls{MLE} approach
each other at a rate of $\mathcal{O}(\nicefrac{1}{n})$. In \citet{Wang:2005},
they show that the asymptotic variational posterior covariance matrix
is ``too small''---it differs from the \gls{MLE} covariance (i.e., the
inverse Fisher information) by a positive-definite matrix.

\subsection{Beyond conditional conjugacy}

We focused on models where the complete conditional is in the
exponential family.  Many models, however, do not enjoy this property.
A simple example is Bayesian logistic regression,
\begin{align*}
  \beta_k &\sim \cN(0, 1),\\
  y_i \g x_i, \beta &\sim \textrm{Bern}(\sigma(\beta^\top x_i)),
\end{align*}
where $\sigma(\cdot)$ is the logistic function.  The posterior
density of the coefficients is not in an exponential family and
we cannot apply the variational inference methods we discussed above.
Specifically, we cannot compute the expectations in the first term of
the \gls{ELBO} in Equation\nobreakspace \textup {(\ref {eq:elbo})} or the coordinate update in
Equation\nobreakspace \textup {(\ref {eq:optimal-qj})}.

Exploring variational methods for such models has been a fruitful
area of research.  An early example is
\citet{jaakkola1997variational,Jaakkola:2000}, who developed a
variational bound tailored to logistic regression. \citet{Blei:2007}
later adapted their idea to nonconjugate topic models, and researchers
have continued to improve the original bound
\citep{khan2010variational,marlin2011piecewise,ermis2014iterative}.
In other work,
\citet{Braun:2010} derived a variational inference algorithm for the
discrete choice model, which also lies outside of the class of
conditionally conjugate models.  They developed a delta method to
approximate the difficult-to-compute expectations.  Finally,
\citet{Wand:2011} use auxiliary variable methods, quadrature, and
mixture approximations to handle a variety of likelihood terms that
fall outside of the exponential family.

More recently, researchers have generalized nonconjugate inference,
seeking recipes that can be used across many models.
\citet{Wang:2013} adapted Laplace approximations and the delta method
to this end, improving inference in nonconjugate generalized linear
models and topic models; this approach is also used by
\citet{Bugbee:2016} for semi-parametric regression.
\citet{Knowles:2011} generalized the
\citet{jaakkola1997variational,Jaakkola:2000} bound in a
message-passing algorithm and \citet{Wand:2014} further simplified and
extended their approach. \citet{Tan:2013,tan2014stochastic}
applied these message-passing methods to
generalized linear mixed models (and also combined them with
\gls{SVI}). \citet{rohde2015} unified many of these algorithmic
developments and provided practical insights into their numerical
implementations.

Finally, there has been a flurry of research on optimizing difficult
variational objectives with \gls{MC} estimates of the gradient.
The idea is to write the gradient of the \gls{ELBO} as an expectation,
compute \gls{MC} estimates of it, and then use stochastic optimization with
repeated \gls{MC} gradients. This first appeared independently in several
papers~\citep{Ji:2010,Nott:2012,Paisley:2012a,Wingate:2013}.  The
newest approaches avoid any model-specific derivations, and are termed
``black box'' inference methods.  As examples,
see~\citet{kingma2013auto,rezende2014stochastic,ranganath2014black,
  Ranganath:2016,
  salimans2014using,titsias2014doubly}, and \citet{tran2016variational}. \citet
  {kucukelbir2016automatic}
leverage these ideas toward an automatic \gls{VI} technique that works
on any model written in the probabilistic programming system Stan
\citep{stan-manual:2015}.  This is a step towards a derivation-free,
easy-to-use \gls{VI} algorithm.

\subsection{Open problems}
\label{sec:open-problems}

There are many open avenues for statistical research in variational
inference.

We focused on optimizing $\kl{q(\bz) || p(\bz \g \bx)}$ as the
variational objective function.  A promising avenue of research is to
develop variational inference methods that optimize other measures,
such as $\alpha$-divergence measures.  As one example, expectation
propagation~\citep{minka2001expectation} is inspired by the \gls{KL}
divergence ``in the other direction,'' between $p(\bz \g \bx)$ and $q(\bz)$.  Other
work has
developed divergences based on lower bounds that are tighter than the
\gls{ELBO}~\citep{Barber:1999,Leisink:2001}. While alternative
divergences may be difficult to optimize, they may give better
approximations~\citep{Minka:2005,opper2005expectation}.

Though it is flexible, the mean-field family makes strong independence
assumptions.  These assumptions help with scalable optimization, but
they limit the expressibility of the variational family.  Further, they
can exacerbate issues with local optima of the objective and
underestimating posterior variances; see Figure\nobreakspace \ref {fig:accuracy}. A
second avenue of research is to develop better approximations while
maintaining efficient optimization.

As we mentioned above, structured variational inference has its roots
in the early days of the method~\citep{Saul:1996a,Barber:1999a}.  More
recently, \citet{hoffman2014structured} use generic structured
variational inference in a stochastic optimization algorithm;
\citet{kucukelbir2016automatic},
\citet{challis2013gaussian},
and \citet{Tan:2016} take advantage of
Gaussian variational families with non-diagonal covariance;
\citet{giordano2015linear} post-process the mean-field parameters to
correct for underestimating the variance; and \citet{Ranganath:2016}
embed the mean-field parameters themselves in a hierarchical model to
induce variational dependencies between latent variables.

The interface between variational inference and \gls{MCMC} remains
relatively unexplored. \citet{Freitas:2001} used fitted variational
distributions as a component of a proposal distribution for
Metropolis-Hastings. \citet{Mimno:2012} and \citet{hoffman2014structured}
studied
\gls{MCMC} as a method of approximating coordinate updates, e.g., to
include structure in the variational family.  \citet{Salimans:2015} propose a
variational approximation to the \gls{MCMC} chain; their method enables an
explicit trade off between computational accuracy and speed.  Understanding how
to combine these two strategies for approximate inference is a ripe area for
future research.  A principled analysis of when to use (and combine) variational
inference and \gls{MCMC} would have both theoretical
and practical impact in the field.

Finally, the statistical properties of variational inference are not
yet well understood, especially in contrast to the wealth of analysis
of \gls{MCMC} techniques.  There has been some progress; see
Section\nobreakspace \ref {sec:theory}.  A final open research problem is to understand
variational inference as an estimator and to understand its
statistical profile relative to the exact posterior.

\appendix

\clearpage
\section{Bayesian Linear Regression with Automatic Relevance\break
Determination}
\label{app:bayes_linear_regression}

Consider a dataset of $\by = y_{1:n} \in \realline^n$ outputs and
$\bx = x_{1:n} \in \realline^{(n \times D)}$ $D$-dimensional inputs, where each
$x_i \in \realline^D$.

A linear regression model assumes a linear relationship between the inputs and
the conditional mean of the output given the inputs.
The latent variable $\beta \in \realline^D$ is a vector of the
regression coefficients.

\Gls{ARD} assigns a separate prior for each regression coefficient; the idea is
to automatically shrink irrelevant coefficients during inference
\citep{mackay1992bayesian,neal2012bayesian,tipping2001sparse,wipf2008new}.
\gls{ARD} works by pairing the prior precision of each regression coefficient
with its own latent variable $\alpha_d$. The hyper-prior on these relevance
variables $\alpha$ encourages small values; this, in turn, selects relevant
regression coefficients.

Here we present a conditionally
conjugate Bayesian linear regression model with an \gls{ARD} prior, based on
\citet{drugowitsch2013variational}.
All Gaussian distributions below follow the precision parameterization.

Define a Gaussian likelihood with precision parameter $\tau$ as
\begin{align*}
  p(\by \mid \beta, \tau \;;\; \bx)
  &=
  \prod_{i=1}^n
  \cN(y_i \mid \beta^\top x_i, \tau).
\end{align*}

\gls{ARD} then posits the following hierarchical prior
\begin{align*}
  p(\beta,\tau \mid \alpha)
  &=
  \cN (\beta \mid 0, \tau \diag(\alpha))
  \;
  \Gam(\tau \mid a_0, b_0),
\end{align*}
where $\alpha$ is a $D$-dimensional relevance variable
\begin{align*}
  p(\alpha)
  &=
  \prod_{d=1}^D
  \Gam(\alpha_d \mid c_0, d_0).
\end{align*}
Here $a_0, b_0, c_0,$ and $d_0$ are fixed hyper-parameters. The latent variable
$\alpha$ determines the relevance of each regression coefficient.

The posterior $p(\beta, \tau, \alpha \mid \by \;;\; \bx)$
is not available in closed form. A simpler model that does not include
$\alpha$ admits a closed form posterior because the normal-gamma distribution
is conjugate to a normal likelihood with unknown mean and precision.

We derive a \gls{CAVI} algorithm for this model. First, factorize the
variational approximation as
\begin{align*}
  q(\beta,\tau,\alpha)
  &=
  q(\beta,\tau) q(\alpha).
\end{align*}
Here we consider $\beta$ and $\tau$ in a single factor.

Begin by applying Equation\nobreakspace \textup {(\ref {eq:optimal-qj})} to identify the optimal form of
$q(\beta,\tau)$ as
\begin{align*}
  \log q(\beta,\tau)
  &=
  \log p(\by \mid \beta, \tau \;;\; \bx)
  +
  \E_{\alpha} [ \log p(\beta,\tau \mid \alpha) ]
  +
  \text{const.}\\
  &=
  \left( \frac{D}{2} + a_0 - 1 + \frac{n}{2} \right) \log \tau \\
  & \quad - \frac{\tau}{2} \Bigg( \beta^\top \left( \E_{\alpha}[\diag\alpha]
  + \sum_i x_i x_i^\top \right) \beta
  + \sum_i y_i^2 - 2 \beta^\top \sum_i x_i y_i + 2 b_0 \Bigg)\\
  & \quad + \text{const.}\\
  &=
  \log \cN(\beta \mid \beta_*, \tau V_*^{-1})
  +
  \log \Gam(\tau \mid a_*, b_*).
\end{align*}
The optimal variational approximation to the regression coefficients and the
precision is thus a normal-gamma with the following parameters:
\begin{align*}
  V_*^{-1} &= \E_{\alpha}[\diag \alpha] + \sum_i x_i x_i^\top, \\
  \beta_* &= V_* \sum_i x_i y_i , \\
  a_* &= a_0 + \frac{n}{2}, \\
  b_* &= b_0 + \frac{1}{2} \left( \sum_i y_i^2 - \beta_*^\top
    V_*^{-1} \beta_* \right).
\end{align*}

Next consider the optimal form of the relevance variables $\alpha$.
Again, apply Equation\nobreakspace \textup {(\ref {eq:optimal-qj})} to identify the optimal form of
$q(\alpha) = \prod_{d=1}^D q(\alpha_d)$ as
\begin{align*}
  \log q(\alpha_d)
  &=
  \E_{\beta,\tau}[\log p(\beta,\tau\mid\alpha_d)]
  +
  \log p(\alpha_d) + \text{const.}\\
  &=
  \left( c_0 - 1 + \frac{D}{2} \right) \log \alpha_d - \alpha_d \left(
    d_0 + \frac{1}{2} \E_{\beta,\tau}[\tau \beta_{d}^2]
  \right) + \text{const.}\\
  &=
  \log \Gam(\alpha_d \mid c_*, d_{*d}).
\end{align*}
The optimal variational approximation to the relevance variable is thus a Gamma
with the following parameters:
\begin{align*}
c_* &= c_0 + \frac{1}{2} , \\
d_{*d} &= d_0 + \frac{1}{2} \E_{\beta, \tau}[\tau \beta_d^2].
\end{align*}
Finally, compute the expectations as
\begin{align*}
  \E_{\alpha}[\diag \alpha]
  &=
  c_* \diag 1/d_{*},\\
  \E_{\beta, \tau}[\tau \beta_d^2]
  &=
  \beta_{*d}^2 a_* / b_* + [V_*]_{d},
\end{align*}
where $[\cdot]_{d}$ indicates the $d$th diagonal entry of a matrix.

Iteratively updating $a_*, b_*, c_*, d_*, V_*^{-1},$ and $\beta_*$ defines
\gls{CAVI} for this model. These quantities also define the \gls{ELBO}; \citet{drugowitsch2013variational} presents the additional algebra that
computes the \gls{ELBO}.

\clearpage
\section{Gaussian Mixture Model of Image Histograms}
\label{app:cavi}

We present a multivariate ($D$-dimensional), diagonal covariance \gls{GMM}.
Denote a dataset of $n$ observations as $\bx = x_{1:n} \in \realline^{(n \times
D)}$, where each $x_i \in \realline^D$. Assume $K$ mixture components.

The cluster assignment latent variables are
$\bz = z_{1:n} \in \realline^{(n \times K)}$
where each $z_i$ is a $K$-indicator vector. The cluster assignments depend on
the mixing vector latent variable $\pi$, which lives in a
$K$-simplex.

The mean latent variables are
$\bmu = \mu_{1:K} \in \realline^{(K \times D)}$, where
each $\mu_k \in \realline^D$, and the precision latent variables are
$\btau = \tau_{1:K} \in \realline^{(K \times D)}$, where each
$\tau_k \in \realline_{>0}^D$.

The joint density of the model factorizes as
\begin{align*}
  p(\bx, \bz, \pi, \bmu, \btau)
  &=
  p(\bx \mid \bz, \bmu, \btau) p(\bz \mid \pi) p(\pi) p(\bmu \mid \btau) p(\btau) .
\end{align*}

The likelihood is Gaussian with precision parameterization
\begin{align*}
  p(\bx \mid \bz, \bmu, \btau)
  &=
  \prod_{i=1}^n \prod_{k=1}^K \left( \prod_{d=1}^D \cN(x_{id} \mid \mu_{kd},
  \tau_{kd}) \right)^{z_{nk}}.
\end{align*}

The marginal over cluster assignments is a categorical distribution,
\begin{align*}
  p(\bz \mid \pi) = \prod_{i=1}^n \prod_{k=1}^K \pi_k^{z_{ik}}.
\end{align*}

The prior over the mixing vector is a Dirichlet distribution with fixed
hyperparameters $a_0$,
\begin{align*}
  p(\pi) = \Dir(\pi \mid a_0 ) = C(a_0) \prod_{k=1}^K \pi_k^{a_0-1}.
\end{align*}

The prior over mean and precision parameters is a normal-gamma distribution with
hyperparameters $m_0$, $b_0$, $\alpha_0$, $\beta_0$,
\begin{align*}
  p(\bmu \mid \btau) p(\btau)
  &=
  \prod_{k=1}^K \prod_{d=1}^D \cN(\mu_{kd} \mid m_0, b_0 \tau_{kd})
  \times
  \prod_{k=1}^K \prod_{d=1}^D \Gam(\tau_{kd} \mid \alpha_0, \beta_0)\\
  &=
  \prod_{k=1}^K \prod_{d=1}^D \cN(\mu_{kd} \mid m_0, b_0 \tau_{kd})
  \:
  \Gam(\tau_{kd} \mid \alpha_0, \beta_0).
\end{align*}

We use the following values for the hyperparameters
\begin{align*}
  a_0 &= \frac{1}{K},\qquad
  m_0 = 0.0,\qquad
  b_0 = 1.0,\qquad
  \alpha_0 = 1.0,\qquad
  \beta_0 = 1.0.
\end{align*}

\citet[Chapter 10.2]{Bishop:2006} derives a \gls{CAVI} algorithm for this
model.

Figure\nobreakspace \ref {fig:code_gmm_diag} presents Stan code that implements this model. Since
Stan does not support discrete latent variables, we marginalize over the
assignment variables.

\begin{figure}[htbp]
\centering
\begin{lstlisting}
data {
  int<lower=0> N; // number of data points in dataset
  int<lower=0> K; // number of mixture components
  int<lower=0> D; // dimension
  vector[D] x[N]; // observations
}

transformed data {
  vector<lower=0>[K] alpha0_vec;
  for (k in 1:K) {           // convert the scalar dirichlet prior 1/K
    alpha0_vec[k] <- 1.0/K;  // to a vector
  }
}

parameters {
  simplex[K] theta;             // mixing proportions
  vector[D] mu[K];              // locations of mixture components
  vector<lower=0>[D] sigma[K];  // standard deviations of mixture components
}

model {
  // priors
  theta ~ dirichlet(alpha0_vec);
  for (k in 1:K) {
      mu[k] ~ normal(0.0, 1.0/sigma[k]);
      sigma[k] ~ inv_gamma(1.0, 1.0);
  }

  // likelihood
  for (n in 1:N) {
    real ps[K];
    for (k in 1:K) {
      ps[k] <- log(theta[k]) + normal_log(x[n], mu[k], sigma[k]);
    }
    increment_log_prob(log_sum_exp(ps));
  }
}
\end{lstlisting}
\caption{Stan code for the \gls{GMM} of image histograms.}
\label{fig:code_gmm_diag}
\end{figure}

\clearpage
\section{Latent Dirichlet Allocation}
\label{app:lda}

Probabilistic topic models are mixed-membership models of text, used to uncover
the latent ``topics'' that run through a collection of documents. Topic models
have become a popular technique for exploratory data analysis of large
collections~\citep{Blei:2012}.

\glsreset{LDA}
\Gls{LDA} \citep{Blei:2003b} is a conditionally conjugate topic model
(Section\nobreakspace \ref {sec:cond-conj}).
It treats documents as containing multiple topics, where a topic is a
distribution over words in a vocabulary.

Following the notation of \citet{Hoffman:2013}, let $K$ be a specific number of
topics and $V$ the size of
the vocabulary. \gls{LDA} defines the following generative process:
\begin{enumerate}
  \item For each topic in $k=1,\ldots,K$,
  \begin{enumerate}
    \item draw a distribution over words $\beta_k \sim \Dir_V(\eta)$.
  \end{enumerate}
  \item For each document in $d=1,\ldots,D$,
  \begin{enumerate}
    \item draw a vector of topic proportions $\theta_d \sim \Dir_K(\alpha)$.
    \item For each word in $n=1,\ldots,N$,
    \begin{enumerate}
      \item draw a topic assignment $z_{dn} \sim \mult(\theta_d)$, then
      \item draw a word $w_{dn} \sim \mult(\beta_{z_{dn}})$.
    \end{enumerate}
  \end{enumerate}
\end{enumerate}
Here $\eta \in \realline_{>0}$ is a fixed parameter of the symmetric
Dirichlet prior on the topics $\beta$, and $\alpha \in \realline_{>0}^K$
are fixed parameters of the Dirichlet prior on the topic proportions for each
document. Similar to the \gls{GMM} example in Section\nobreakspace \ref {sec:mog}, we encode
categorical variables as indicator vectors. Thus $z_{dn}$ is a $K$-vector where
$z_{dn}^k = 1$ indicates the $n$th word in document $d$ is assigned to the
$k$th topic. Similarly, $w_{dn}$ is a $V$-vector where $w_{dn}^v = 1$ indicates
that the $n$th word in document $d$ is the $v$th word in the vocabulary. We
occasionally abuse these indicator vectors as indices---for example, if
$z_{dn}^k=1$, then $\beta_{z_{dn}}$ is the $k$th topic, denoted by $\beta_k$.

The posterior $p(\beta, \theta, z \mid w)$
is not available in closed form. While the topic
assignments
$z$ and their proportions $\theta$ enjoy a conjugate relationship, the
introduction of the topics $\beta$ renders this posterior analytically
intractable.

We derive a \gls{CAVI} algorithm for this model, based on \citet{Hoffman:2013}.
Posit a mean-field variational family
\begin{align*}
  q(\beta, \theta, z)
  &=
  \prod_{k=1}^K q(\beta_k \;;\; \lambda_k)
  \prod_{d=1}^D
  \left(
  q(\theta_{d} \;;\; \gamma_d)
  \prod_{n=1}^{N}
  q(z_{dn} \;;\; \phi_{dn})
  \right).
\end{align*}
Since \gls{LDA} is a conditionally conjugate model, we can directly identify the
family of each factor (Section\nobreakspace \ref {sec:cond-conj}).

Begin with the complete conditional of the topic assignment. This is a
multinomial,
\begin{align*}
  p(z_{dn} = k \mid \theta_d, \beta, w_{dn})
  &\propto
  \exp
  \left(
  \log\theta_{dk}
  +
  \log\beta_{k,{w_{dn}}}
  \right).
\end{align*}
The variational approximation to the topic assignments is also a
multinomial distribution, with parameters $\phi_{dn}$.

Follow with the complete conditional of the topic proportions. This is
a $K$-dimensional Dirichlet
\begin{align*}
  p(\theta_d \mid z_d)
  &=
  \Dir_K
  \left(
  \alpha + \sum_{n=1}^N z_{dn}.
  \right)
\end{align*}
The variational approximation to the topic
proportions is also a $K$-dimensional Dirichlet with parameters $\gamma_d$.

End with the complete conditional of the topics. This is a $V$-dimensional
Dirichlet
\begin{align*}
  p(\beta_k \mid z, w)
  &=
  \Dir_V
  \left(
  \eta
  +
  \sum_{d=1}^{D}
  \sum_{n=1}^{N}
  z_{dn}^k
  w_{dn}
  \right).
\end{align*}
In words,
the $v$th element of the $k$th topic is a Dirichlet with parameter equal to the
sum of the fixed scalar $\eta$ and the number of times term $v$ (denoted by $w_
{dn}$) was
assigned to topic $k$ (denoted by $z_{dn}^k$). The variational
approximation to the
topic proportions is a $V$-dimensional Dirichlet with parameters
$\lambda_k$.

The \gls{CAVI} updates for the topic assignment and topic proportions require
iterating over the following for each word within each document until
convergence:
\begin{align}
  \phi_{dn}^k
  &\propto
  \exp
  \left(
  \EE{\log\theta_{dk}}
  +
  \EE{\log\beta_{k,{w_{dn}}}}
  \right) \nonumber \\
  &\propto
  \exp
  \left(
  \Psi(\gamma_{dk})
  +
  \Psi(\lambda_{k,{w_{dn}}})
  -
  \Psi\left(\sum_v\lambda_{kv}\right)
  \right) \label{eq:lda_phi} \\
  \gamma_{d}
  &=
  \alpha + \sum_{n=1}^{N}\phi_{dn} \label{eq:lda_gamma}
\end{align}
This is a direct
application of Equation\nobreakspace \textup {(\ref {eq:local-update})} to the complete conditionals above.

The updates for $\phi$ and $\gamma$ depend on the variational
parameters for the topics $\lambda$. The update for the topics, in
turn, depends on the variational parameters for the topic proportions. That
update is
\begin{align}
  \lambda_k
  &=
  \eta
  +
  \sum_{d=1}^{D}
  \sum_{n=1}^{N}
  \phi_{dn}^k
  w_{dn}. \label{eq:lda_lambda}
\end{align}
This update only depends on the
variational parameter for the topic assignments $\phi_{dn}$.

Algorithm\nobreakspace \ref {alg:cavi_lda} presents the full \gls{CAVI} algorithm for \gls{LDA}. A
similar computation defines the \gls{ELBO} for \gls{LDA}; \citet{Hoffman:2013}
present the additional algebra for the \gls{ELBO}.

\begin{algorithm}[!hbtp]
\setstretch{1.25}
\KwIn{\gls{LDA} model and a set of words in documents $w$.}
\KwOut{Variational parameters $\lambda, \gamma, \phi$.}
\textbf{Initialize:} Variational parameters $\lambda, \gamma$ randomly.\\
\While{the \gls{ELBO} has not converged}{
  \Repeat{$\phi$ and $\gamma$ have converged}
  {
    \For{each document $d$}
    {
      \For{each word $n$}
      {
        Compute updates to $\phi$ and $\gamma$ via
        Equations\nobreakspace \textup {(\ref {eq:lda_phi})} and\nobreakspace  \textup {(\ref {eq:lda_gamma})}.
      }
    }
  }
  Compute update to $\lambda$ via Equation\nobreakspace \textup {(\ref {eq:lda_lambda})}.
}
\caption{\gls{CAVI} for \gls{LDA}}
\label{alg:cavi_lda}
\end{algorithm}

\clearpage
\bibliographystyle{apa}
\bibliography{bib}

\end{document}